\documentclass[useAMS,usenatbib]{mn2e}
\usepackage{amssymb}
\usepackage{amsmath}
\usepackage[dvips]{graphicx}
\usepackage{natbib}
\usepackage{lscape}

\title[Broadband modelling of magnetars in SGRBs]{Broadband modelling of short gamma-ray bursts with energy injection from magnetar spin-down and its implications for radio detectability}
\author[B. P. Gompertz et al.]
{\parbox{\textwidth}{B. P. Gompertz$^{1}$\thanks{E-mail: bpg6@le.ac.uk}, A. J. van der Horst$^{2}$, P. T. O'Brien$^{1}$, G. A. Wynn$^{1}$ and K. Wiersema$^{1}$}\vspace{0.4cm}\\
\parbox{\textwidth}{$^{1}$Department of Physics and Astronomy, University of Leicester, Leicester, LE1 7RH, UK\\
$^{2}$Anton Pannekoek Institute, University of Amsterdam, Science Park 904, 1098 XH Amsterdam, The Netherlands}}

\begin{document}

\date{Accepted:}

\pagerange{\pageref{firstpage}--\pageref{lastpage}} \pubyear{????}

\maketitle

\label{firstpage}

\begin{abstract}
The magnetar model has been proposed to explain the apparent energy injection in the X-ray light curves of short gamma-ray bursts (SGRBs), but its implications across the full broadband spectrum are not well explored. We investigate the broadband modelling of four SGRBs with evidence for energy injection in their X-ray light curves, applying a physically motivated model in which a newly formed magnetar injects energy into a forward shock as it loses angular momentum along open field lines. By performing an order of magnitude search for the underlying physical parameters in the blast wave, we constrain the characteristic break frequencies of the synchrotron spectrum against their manifestations in the available multi-wavelength observations for each burst. The application of the magnetar energy injection profile restricts the successful matches to a limited family of models that are self-consistent within the magnetic dipole spin-down framework. We produce synthetic light curves that describe how the radio signatures of these SGRBs ought to have looked given the restrictions imposed by the available data, and discuss the detectability of these signatures with present-day and near-future radio telescopes. Our results show that both the Atacama Large Millimetre Array and the upgraded Very Large Array are now sensitive enough to detect the radio signature within two weeks of trigger in most SGRBs, assuming our sample is representative of the population as a whole. We also find that the upcoming Square Kilometre Array will be sensitive to depths greater than those of our lower limit predictions.
\end{abstract}

\section{Introduction}\label{sec:intro}

Gamma-ray bursts (GRBs) are extreme outbursts of electromagnetic radiation, releasing energies of the order of $10^{48}-10^{52}$~erg in a relativistic outflow, when collimation of this outflow is accounted for \citep[e.g.,][]{Cenko11}. They are divided into two classes: long and short GRBs (LGRBs and SGRBs, respectively), sitting at either side of a $T_{90} \sim 2$~s divide \citep{Kouveliotou93}, where $T_{90}$ is the time in which the cumulative counts increase from 5 to 95 per cent of the background level. SGRBs typically emit a short ($T_{90} < 2$~s) prompt spike of gamma radiation which decays away into an X-ray afterglow, but $\sim 50$ per cent of cases \citep[e.g.][]{Rowlinson13} exhibit a flat X-ray plateau which lasts for around a thousand seconds. In some SGRBs, the initial decay is interrupted by a prolonged, softer, rebrightening in the high energy light curves. This extended emission (EE) was discovered in $\sim1/3$ of SGRBs \citep{Norris06}. It usually begins $\lesssim10$~s after the trigger, and while it often has a lower luminosity than the prompt emission, it can last for a few hundred seconds, implying that the total energy contained is comparable \citep{Perley09}. Those bursts that were believed to exhibit EE were catalogued by \citet{Norris10}, and that sample was updated and expanded in \citet{Gompertz13}.

One of the leading models for SGRBs is the merger of two compact objects: some combination of black holes, neutron stars (NS), and/or white dwarfs \citep{Paczynski86,Rosswog03,Belczynski06,Chapman07}. Two possible post-merger remnants have been proposed as the central engine of SGRBs: a black hole \citep{Woosley93,Kumar08}, or a rapidly-rotating, highly-magnetized NS known as a magnetar \citep{Gao06,Metzger08,Bucciantini12,Cheng14,Lu14,Metzger14}. In this paper, we consider the magnetar case. These magnetars can have dipole fields of the order of $10^{15}$~G at birth \citep{Duncan92,Thompson95,Kouveliotou99,Esposito10} and spin at periods of around $1$~ms \citep{Lattimer04,Metzger11,Giacomazzo13b}. In this context, the prompt emission is considered to be due to relativistic jets powered by the initial merger, and the X-ray plateau seen in $\sim 50$ per cent of bursts \citep{Rowlinson13} is the result of energy injection into the radiating shock via magnetic dipole spin-down. The luminosity and duration of the X-ray plateau is then just a function of the magnetar's dipole field strength and spin period \citep{Zhang01}, and the observed anticorrelation between them \citep{Dainotti08} has been suggested as a natural prediction of the model \citep{Rowlinson14}. The magnetar model has been applied to LGRBs \citep{Zhang01,Troja07,Lyons10,Dall'Osso11,Metzger11,Bernardini12}, SGRBs \citep{Fan06,Rowlinson10,Rowlinson13} and EE GRBs \citep{Metzger08,Bucciantini12,Gompertz13,Gompertz14}.

Most applications of the magnetar model on GRB light curves have focused on the X-ray regime. In these efforts, bolometric light curves are constructed based on the X-ray and soft gamma-ray light curves, and they are modelled with a time-dependent luminosity that reflects the magnetic dipole spin-down of a rapidly rotating magnetar \citep{Zhang01}. It is typically assumed that the spin-down luminosity evolution is converted to an observed light curve evolution with a certain efficiency factor. This method does not provide information on light curves at lower frequencies, although extending the spectrum to the optical bands has been attempted \citep[e.g.][]{Rowlinson13}. In this paper, we aim to perform broadband modelling of SGRB afterglows, using the available light curves at X-ray, optical, and in some cases radio bands, all within the magnetar model. Instead of assuming a direct conversion of magnetic dipole spin-down into light curves at different frequencies, we treat this spin-down as energy injection into the shocks at the front of the relativistic outflow. These shocks emit broadband synchrotron radiation, and we calculate the light curves across the spectrum as is typically done for GRB afterglows \citep[e.g.][]{Sari98,Wijers99} but with a time-varying energy term governed by magnetic dipole spin-down. We apply this model to four SGRBs to illustrate the method, and we give ranges for the physical parameters of the shocks and the magnetar for these four sources. We also discuss the detectability of the range of light curves our models predict for radio emission in the context of current and next generation radio telescopes.

In Section~\ref{sec:data} we describe the GRBs in our sample and the broadband data we collected for all four of them. The model and methods we used to construct the broadband light curves are given in Section~\ref{sec:model}. The results of modelling the light curves are shown in Section~\ref{sec:results}, followed by a discussion and conclusions in Sections~\ref{sec:discussion}--\ref{sec:conclusions}.

\section{Data Sample}\label{sec:data}

We collected a sample of four SGRBs with good X-ray observations exhibiting a clear plateau, for which there were contemporaneous optical observations in at least one filter and an identified redshift. Radio observations were not essential, but were a welcome bonus. Our sample represents those SGRBs with the best data availability to test the analysis and introduce the model, but is not an exhaustive list of all SGRBs that satisfy the selection criteria. The classification of GRB 060614 is uncertain \citep{Gehrels06}, but we include it as an EE GRB here due to the lack of an associated supernova to deep limits \citep{DellaValle06,GalYam06} as would be expected for an LGRB. Analysis by \citet{Zhang07b} also suggests that this burst is linked to the short class. The X-ray data used here were taken by the \emph{Swift} X-Ray Telescope (XRT; \citealt{Burrows05}) and were downloaded from the UK \emph{Swift} Science Data Centre (UKSSDC) archives \citep{Evans07,Evans09}. We used the $0.3$ -- $10$ keV flux light curves, corrected for absorption using a ratio of (counts-to-flux unabsorbed)/(counts-to-flux observed). Details of the data reduction process can be found in \citet{Evans07,Evans09}. The $0.3$ -- $10$ keV flux light curves were then compressed into flux density light curves at $1.73$ keV (the bandpass logarithmic mid-point) using the equations in Appendix~\ref{sec:equations}.

References for the ultraviolet (UV), optical, infrared (IR), and radio data that were used are shown in Table~\ref{tab:datatable}. Galactic extinction correction was done using the values in \citet{Schlafly11}, even in cases where the original data were corrected using the \citet{Schlegel98} values. In most cases, we neglect the effect of intrinsic absorption due to a lack of high-quality near-IR to optical data with which to constrain it. The single exception is GRB 130603B, which was corrected with $A_v = 0.86 \pm 0.15$~mag and a Small Magellanic Cloud (SMC) extinction law \citep{deUgartePostigo14}. Conflicting values were also derived for the intrinsic absorption in GRB 060614, with \citet{DellaValle06} finding $A_v = 0.08$~mag and \citet{Covino13} finding $A_v = 0.74^{+0.20}_{-0.17}$~mag or $A_v = 0.62 \pm 0.06$~mag, depending on whether an X-ray prior was used. Both studies find an SMC extinction law. The discrepancy comes from the use of a spectral break between the optical and X-ray frequencies, and highlights how the applied model can influence the derived intrinsic absorption. We neglect the intrinsic absorption in this case in an attempt to make our results as general as possible. We also note that \citet{Fong14b} derived $A_v = 0.5$ for GRB 070714B, but we do not include it because it was derived by comparison of the optical and X-ray bands, and is therefore dependent on the presence of a spectral break between them.

\begin{table}
\begin{center}
\begin{tabular}{lccll}
\hline
GRB & $\Gamma$ & $z$ & $E(B-V)$ & Reference \\
\hline
051221A & $1.95^{+0.18}_{-0.17}$ & $0.5465^a$ & $0.069$ & [1] \\
060614 & $1.78^{+0.08}_{-0.08}$ & $0.1254^b$ & $0.019$ & [2,3,4] \\
070714B & $1.76^{+0.28}_{-0.24}$ & $0.9224^c$ & $0.141$ & [5] (A,B,C,D) \\
130603B & $1.98^{+0.15}_{-0.14}$ & $0.356^d$ & $0.02$ & [6,7,8] (E) \\
\hline
\end{tabular}
\caption{UV, optical, IR and radio data used. Photon indices, $\Gamma$ are for the X-ray data, and come from the UKSSDC spectrum repository \citep{Evans07,Evans09} which gives $90$ per cent confidence interval errors. $E(B-V)$ values are from \citet{Schlafly11}. \newline References (redshift): $^a$\citet{Soderberg06}; $^b$\citet{GalYam06}; $^c$\citet{Graham09}; $^d$\citet{Thone13}. \newline References (refereed): [1] - \citet{Soderberg06}; [2] - \citet{DellaValle06}; [3] - \citet{GalYam06}; [4] - \citet{Mangano07}; [5] - \citet{Graham09}; [6] - \citet{Tanvir13}; [7] - \citet{deUgartePostigo14}; [8] - \citet{Fong14}. \newline References (GCN circulars): (A) - \citet{Chandra07}; (B) - \citet{Landsman07}; (C) - \citet{Perley07}; (D) - \citet{Weaver07}; (E) - \citet{dePasquale13}. \label{tab:datatable}}
\end{center}
\end{table}

\section{Model}\label{sec:model}

The central engine in our model is a magnetar, formed as a product of the merger of two NS. The merger drives a relativistic outflow, which expands with time. Internal processes such as shocks between expanding shells of ejecta \citep{Goodman86,Paczynski86} or magnetic turbulence \citep{Zhang11} convert some of the kinetic energy of the blast wave into electromagnetic radiation, which is observed as the SGRB prompt emission. The blast wave sweeps up ambient particles as it expands into the interstellar medium (ISM), and eventually starts to slow down once it has accumulated sufficient mass. This deceleration radius, $R_{\rm dec}$, marks the outer boundary for emission processes to be considered `internal'. The interaction between the blast wave and the ISM forms a strong shock at the head of the ejecta, and a synchrotron emission spectrum is set up by the action of electrons traversing the shock front. This is the emission site of the afterglow. For simplicity, we consider here an adiabatic expansion, and invoke a homogeneous ambient medium, as is expected in the vicinity of an NS binary. However, for LGRBs or binaries with pulsar winds the local density profile can be different \citep[see e.g.][]{Starling08,Curran09}.

\subsection{Synchrotron emission}

The relativistic blast wave accelerates electrons, which subsequently radiate synchrotron emission in accordance with their respective Lorentz factors, which are assumed to have a power-law distribution. There are two distinct emission regimes, dubbed `fast cooling', in which the cooling time-scale of the electrons is less than the lifetime of the source, and `slow cooling', in which the majority of electrons cool on longer timescales than the source lifetime. \citep{Sari98,Wijers99}.

The synchrotron spectrum has three characteristic break frequencies: the peak frequency, $\nu_{\rm m}$; the cooling frequency, $\nu_{\rm c}$; and the self-absorption frequency, $\nu_{\rm a}$, at which the medium changes from being optically thin to being optically thick. These breaks are not static in time, but change and evolve with the hydrodynamical expansion of the blast wave. Their position and evolution determines the phenomenology of the corresponding light curve at a given observational band \citep{Sari98,Wijers99}. The breaks and peak flux ($F_{\nu,\rm max}$) are governed by the energy contained in the blast wave and three other physical parameters: $\epsilon_{\rm e}$, the fraction of energy contained in the emitting electrons, $\epsilon_B$, the fraction of energy contained in the magnetic field, and $n_0$, the number density (in cm$^{-3}$) of particles in the ambient medium.

In addition, the breaks' behaviour is affected by the dynamical state of the blast wave, which can be in the relativistic, jet-spreading, or non-relativistic phase. The jet spreading phase occurs when $\theta_0 \approx \gamma^{-1}$, where $\theta_0$ is the opening angle of the collimated jet, and $\gamma$ is the bulk Lorentz factor of the blast wave. The observer begins to `notice' the edge of the jet as it expands, and as $\gamma$ drops the jet spreads sideways \citep{vanEerten12}. As the shock becomes almost spherical, it becomes non-relativistic at a time $t_{\rm NR}$ which can be approximated by \citep{vanEerten12}
\begin{equation}
t_{\rm NR} = 1100\bigg(\frac{E_{\rm iso}}{10^{53}n_0}\bigg)^{1/3} \mbox{ d.}
\end{equation}
These three dynamical phases each have their own hydrodynamical evolution, and hence the time dependences of the synchrotron break frequencies also vary. Values used for the synchrotron spectrum and its evolution in this paper come from chapter 2 of \citet{vanderHorst07}. In our analysis, we deal only with the forward shock emission.

\subsection{Energy injection}

The magnetar formed by the merger is initially rapidly spinning, with a spin period of the order of $1$~ms. After birth, it loses angular momentum in the form of magnetic dipole spin-down \citep{Zhang01}, resulting in energy being injected into the outflow and the forward shock for a sustained period, typically of the order of $1000$~s. This was investigated for LGRBs by \citet{Dall'Osso11}. The total energy injected into the shock at a time $t$ after merger is given by (cf. \citealt{Zhang01})
\begin{equation}
E_{\rm d}(t)=\frac{\eta L_0t}{(1+t/T_{\rm em})^2}.
\label{eq:dipole}
\end{equation}
The parameter $\eta$ accounts for our ignorance in the efficiency of the transfer of energy from the dipole to the forward shock, both in terms of radiative losses and beaming factor. $L_0$ is the luminosity of the dipole plateau in erg~s$^{-1}$ and $T_{\rm em}$ is the point at which the plateau turns over, known as the characteristic spin-down time-scale. $L_0$ and $T_{\rm em}$ are both derived from the underlying physical parameters of the magnetar:
\begin{align}
L_{0,49} =& \hspace{0.1cm} B^2_{\rm p,15}P^{-4}_{0,-3}R^6_6 \\
T_{\rm em,3}=& \hspace{0.1cm} 2.05I_{45}B^{-2}_{\rm p,15}P^2_{0,-3}R^{-6}_6,
\end{align}
where $L_{0,49}$ is $L_0$ in units of $10^{49}$~erg~s$^{-1}$ and $T_{\rm em,3}$ is $T_{\rm em}$ in units of $10^3$~s. $I_{45}$ is the moment of inertia in units of $10^{45}$~g~cm$^2$, and is $\sim$ 1 (2) for a $1.4$ ($2.1$) $M_{\odot}$ NS. $R_6$ is the NS radius in $10^6$~cm, $P_{0,-3}$ is the spin period in ms and $B_{\rm p,15}$ is the dipole field strength in units of $10^{15}$~G. We set the NS radius $R_6 = 1$, since this is consistent with most equations of state \citep{Lattimer04}. These relations place limits on the values of $L_0$ for a given $T_{\rm em}$, principally through the break-up spin period for an NS (e.g. $P \geq 0.66$ ms for a $2.1M_{\odot}$ NS; \citealt{Lattimer04}). The upper limit placed on $L_0$ by $P$ is given by
\begin{equation}
L_{0,49}\leq2.05I_{45}T^{-1}_{\rm em,3}P^{-2}_{\rm lim,-3}
\label{eq:limit}
\end{equation}
because $T_{\rm em}$ is a fixed quantity for a given GRB.

The two EE bursts in our sample, GRB 060614 and GRB 070714B, are likely to also inject energy into the shock during the EE phase, although without a clear model for what EE is, it is difficult to say how much. To represent EE, we use the energy profile from \citet{Gompertz14}, who used a magnetic propeller to describe the emission feature. These magnetic propellers accelerate infalling material to super-Keplerian velocities, ejecting it from the system at relativistic speeds, where it subsequently shocks to produce electromagnetic radiation. The exact physics behind these models is largely irrelevant for our needs, but the accurate luminosity profile provides a convenient way to introduce EE energy injection to the system. The total energy in the forward shock at a time t is then given by
\begin{equation}
E_{\rm FS}(t)=E_{k}+E_{\rm EE}(t)+E_{\rm d}(t).
\end{equation}
Here, $E_k$ represents the impulsive energy of the blast wave, and is tied to the prompt emission isotropic equivalent energy $E_{\gamma,\rm iso}$ through a prefactor accounting for beaming and efficiency. $E_{\rm EE}$ is the energy injected during EE, representing the luminosity profile from \citet{Gompertz14} multiplied by another prefactor $\kappa$, again to account for beaming and efficiency. $E_{\rm d}$ is the energy injected by dipole spin-down, given by Equation~\ref{eq:dipole}. These energies are varied to obtain fits to the data, and the physical implications that the obtained values have for the central engine are discussed in Section~\ref{sec:discussion}.

\subsection{X-ray and optical fitting}

To perform least-squares fitting for broadband GRB afterglows, one normally requires well-sampled light curves in the X-ray and optical bands, as well as at least two radio bands. Without radio observations, it is very difficult to locate $\nu_{\rm a}$, since this break is normally found at radio frequencies, and $\nu_{\rm m}$ and $F_{\nu,\rm max}$ can only be constrained as a combination, rather than individually. Additionally, if $\nu_{\rm c}$ lies above the X-ray frequency then it too becomes poorly constrained. Because of this, large degeneracies can occur where the observed X-ray and optical light curves give combinations of $\nu_{\rm m}$ and $F_{\nu,\rm max}$ that can be recreated by many different physical parameter values, each having very different implications for the positions of $\nu_{\rm c}$ and $\nu_{\rm a}$. Thus, any fitting can result in parameter uncertainties spanning several orders of magnitude. For our sample, the available data consist of a well-sampled X-ray light curve, as well as a sparsely sampled optical light curve (sometimes in multiple bands) and just one or two radio observations or limits at best per burst. This is insufficient for fitting in the traditional way, so we conduct an order of magnitude search of the parameter space within reasonable parameter limits.

Synthetic light curves are created through a combination of nine free parameters. Three are well constrained by the data: the characteristic spin-down time-scale $T_{\rm em}$, the jet break time $t_{\rm jb}$, and the power-law index of the electron Lorentz factor distribution $p$. $p$ is the most constrained; this parameter sets the spectral slope, so the simultaneous goodness-of-fit to both the X-ray and R-band data is very sensitive to its value (with a small mitigation for the position of the cooling break: $\beta = \frac{p-1}{2}$ for $\nu_{\rm m} < \nu < \nu_{\rm c}$; $\beta = \frac{p}{2}$ for $\nu_{\rm c} < \nu$). $p$ also sets the temporal decay of the light curves, adding further constraint to its value. Because of these strong constraints, we use a single value of $p$, obtained by simultaneous model fitting to both the X-ray and optical light curves, as well as the late-time temporal decay in the post-plateau region.

Once this value is obtained, the next most constrained parameter is $T_{\rm em}$, which determines the time at which the flat plateau region transitions into the late-time temporal decay. There is some degeneracy between the temporal slope of the decay (controlled by $p$) and the time at which transition occurs (controlled by $T_{\rm em}$), particularly in cases where data in this region is sparse, but the extra constraint on $p$ from the spectral slope requirements ensures that a single value can be used for both parameters; values of $p$ outside of a fairly small range are unable to provide simultaneous fits to the X-ray and optical light curves. In cases where the late temporal decay is too steep at both X-ray and optical frequencies for any reasonable combination of $p$ and $T_{\rm em}$ to reproduce, a jet break is used, implemented as a smooth achromatic break at a time $t_{\rm jb}$. Where no jet break was required at all, we tested models assuming no jet break and ones assuming the earliest jet break allowed by the data to produce the full range of possible fluxes. The single-value model parameters are listed in Table~\ref{tab:parameters}.

\begin{table}
\begin{center}
\begin{tabular}{lccc}
\hline
GRB & $p$ & $T_{\rm em}$ & $t_{\rm jb}$ \\
& & (s) & (d) \\
\hline
051221A & $2.4$ & $8.0 \times 10^3$ & $\geq 4.0$ \\
060614 & $2.6$ & $2.5 \times 10^4$ & $1.10$ \\
070714B & $2.9$ & $2.0 \times 10^3$ & $\geq 0.7$ \\
130603B & $2.5$ & $8.0 \times 10^2$ & $0.35$ \\
\hline
\end{tabular}
\caption{The single-value free parameters for each burst, selected by data constraints. \label{tab:parameters}}
\end{center}
\end{table}

The remaining six parameters are less constrained. They are $\epsilon_{\rm e}$, $\epsilon_B$, $n_0$, $L$ (where $L = \eta L_0$), $\kappa$ and $E_k$. We apply constraints to the range of allowed values for these parameters. $\epsilon_B$ has been found to be as low as $10^{-8}$ \citep{BarniolDuran14,Santana14} and as a fraction can be as high as $1$. In practice, $\epsilon_{\rm e}$ tends towards higher values than $\epsilon_B$. We set an upper limit of $1$, noting that $\epsilon_{\rm e}$ actually refers to the electron population that is emitting synchrotron radiation, rather than the electron population as a whole, and set a lower limit of $10^{-3}$ \citep{Kumar00b}. $n_0$ is limited between $10^{-5}$ and $100$~cm$^{-3}$, in line with what has been found in these sources \citep{Cenko11}. The upper limit of $L$ is set by the argument in Equation~\ref{eq:limit}, and values of this parameter below $\sim 10^{47}$~erg~s$^{-1}$ are never energetic enough to match the data, so we set the lower limit as $10^{47}$~erg~s$^{-1}$. Within these limits for $L$, we find that EE ceases to have any influence on the light curve if $\kappa \lesssim 10^{-2}$. If EE is isotropic, and the observed luminosity is only 1 per cent of the true energy (i.e. the conversion efficiency of kinetic to potential energy in the internal shocks is 1 per cent), then the energy delivered to the synchrotron shock front could be up to 100 times higher than observed in the light curve. In practice, however, the emission is (a) unlikely to be fully isotropic, (b) likely to shock more efficiently than 1 per cent, and (c) certain to be less than 100 per cent efficient at delivering its energy to the synchrotron shock front. For these reasons, we set the upper limit of $\kappa$ at a still fairly generous factor of 10. Finally, we limit the energy in the shock from prompt emission to $10^{48}$~erg $ < E_k < 10^{52}$~erg. The arguments for these limits are identical to those used for $\kappa$, except that the prompt emission is known to be beamed \citep{Sari99,Frail01} so the upper limit is lower, and because the injected energy at early times is negligible, $E_k$ dominates the early light curve so the lower limit can be much less energetic before its influence vanishes. These limits are summarized in Table~\ref{tab:magsearch}.

\begin{table}
\begin{center}
\begin{tabular}{lcc}
\hline
Parameter & Minimum & Maximum \\
\hline
$\epsilon_{\rm e}$ & $10^{-3}$ & $1$ \\
$\epsilon_B$ & $10^{-8}$ & $1$ \\
$n_0$ (cm$^{-3}$) & $10^{-5}$ & $100$ \\
$L$ (erg~s$^{-1}$) & $10^{47}$ & $10^{49a}$ \\
$\kappa$ & $10^{-2}$ & $10$ \\
$E_k$ (erg) & $10^{48}$ & $10^{52}$ \\
\hline
\end{tabular}
\caption{Limits on parameters used in the order of magnitude parameter space search. $^a10^{49}$~ergs~s$^{-1}$ is typical, but the real value depends on Equation~\ref{eq:limit} \label{tab:magsearch}}
\end{center}
\end{table}

Each combination of parameters creates a synthetic light curve, and the match to the data is assessed by calculating the $\chi^2$ value for the X-ray observations, as well as observations in the R-band since this is always the best sampled optical light curve. The $\chi^2$ values for the two light curves are assessed separately to avoid a situation where an excellent fit to the X-rays but a poor fit to the optical is indistinguishable from a good fit to both, since the statistics will be dominated by the much better sampled X-ray light curve. Upper limits are not included in the $\chi^2$ calculations, but were subsequently inspected for violations (see Section~\ref{sec:results}). Since there are often fewer R-band data points than free parameters, we are not able to calculate the reduced $\chi^2$ for the individual bands, but do calculate the overall reduced $\chi^2$ by summing the $\chi^2$ contribution and dividing by the combined degrees of freedom. The X-ray band $\chi^2$ is obtained for data points in the X-ray plateau and later, excluding the preceding steep decay. This region is believed to be due the curvature effect \citep{Kumar00}.

\begin{figure*}
\begin{center}
\includegraphics[width=8.8cm]{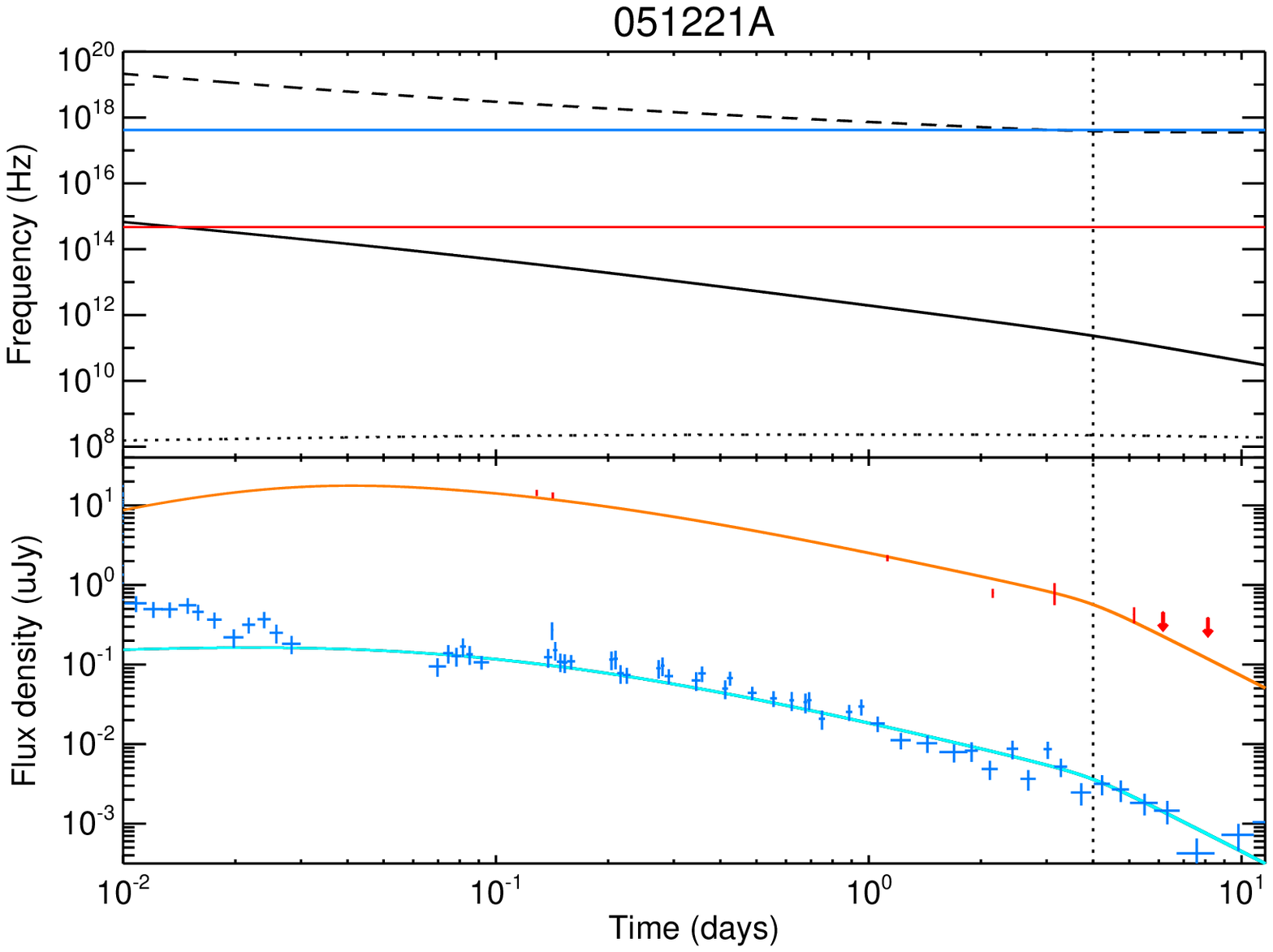}
\includegraphics[width=8.8cm]{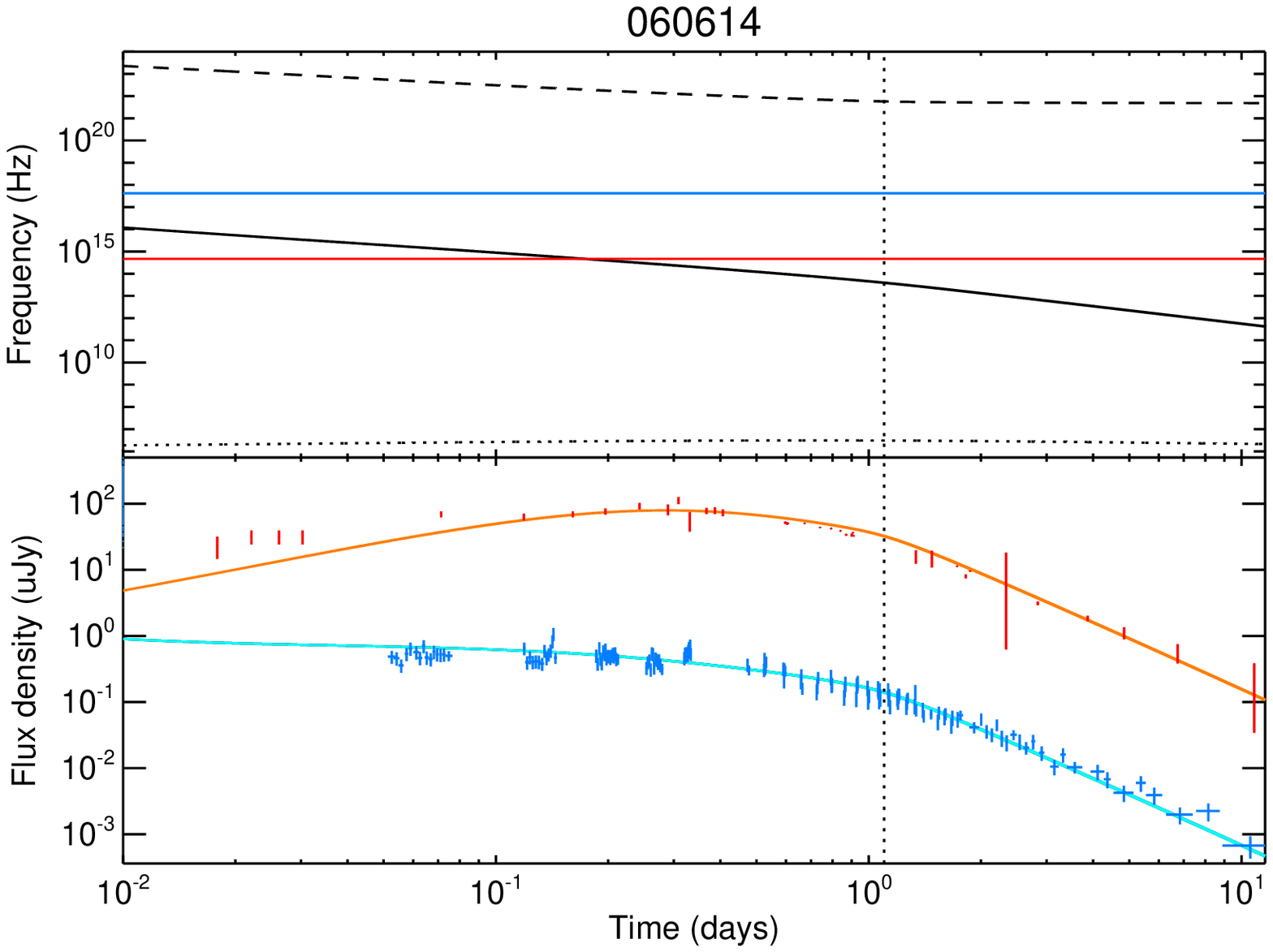}
\includegraphics[width=8.8cm]{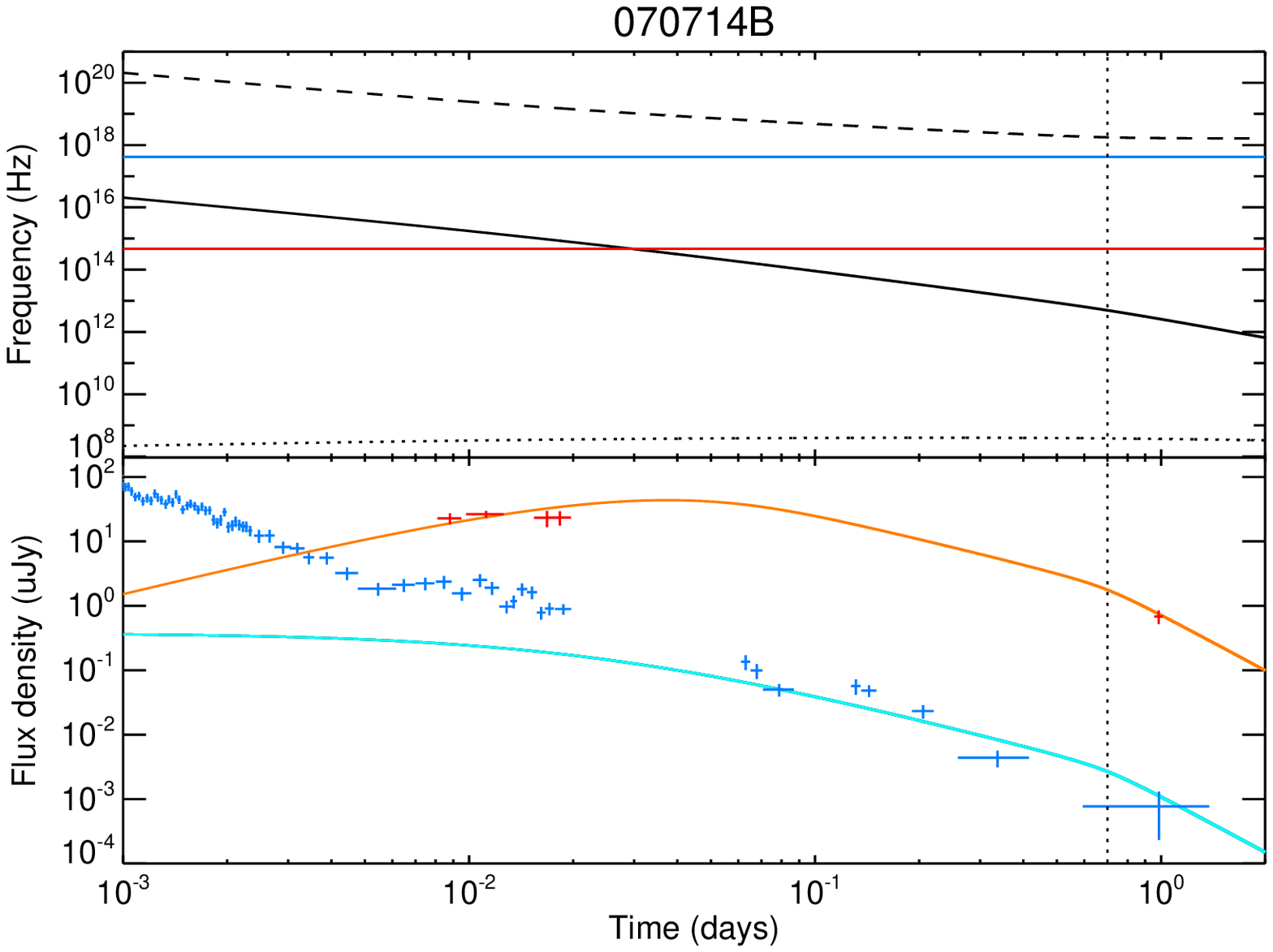}
\includegraphics[width=8.8cm]{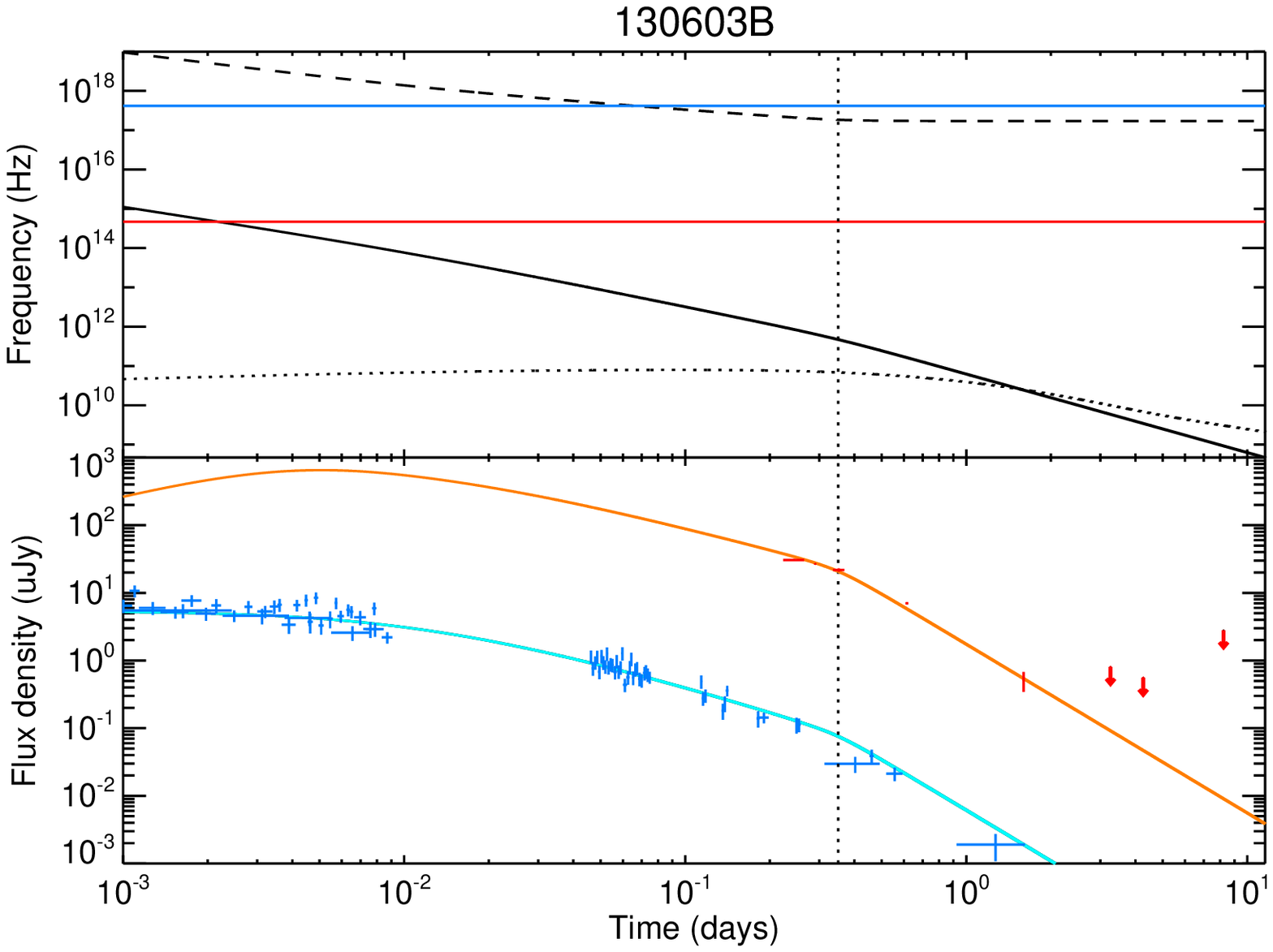}
\end{center}
\caption{Example fits to the X-ray and R-band light curves of the four GRBs in our sample. Top panels: temporal evolution of the three spectral breaks. The black dashed, solid, and dotted lines are the cooling, peak, and self-absorption breaks, respectively. The horizontal blue (red) line marks the X-ray (R-band) frequency for reference. The vertical black dotted line denotes a jet break. Bottom panels: light curves showing the model fit line to the X-ray (blue) and R-band (red) data points. The goodness-of-fit in the X-ray band is only assessed for the plateau data and later, i.e. all data in 060614 and 130603B, and data at times later than $5\times10^{-2}$~d in 051221A and 070714B. The rebrightening at around $10^{-2}$~d in GRB 070714B is interpreted as an X-ray flare. \label{fig:example}}
\end{figure*}

\begin{table*}
\begin{center}
\begin{tabular}{lccccccc}
\hline
GRB & Reduced & $\epsilon_{\rm e}$ & $\epsilon_B$ & $n_0$ & $L$ & $E_k$ & $\kappa$ \\
& $\chi^2$ limit & & & (cm$^{-3}$) & (erg~s$^{-1}$) & erg & \\
\hline
051221A & 2.8 & $0.1$--$1$ & $10^{-4}$--$10^{-1}$ & $10^{-4}$--$10^1$ & $10^{47}$--$10^{48}$ & $10^{48}$--$10^{51}$ & -- \\
060614 & 15 & $0.1$--$1$ & $10^{-7}$--$10^{-3}$ & $10^{-5}$--$10^{2}$ & $10^{48}$ & $10^{48}$--$10^{50}$ & $10^{-2}$--$10^{-1}$ \\
070714B & 10 & $0.1$--$1$ & $10^{-6}$--$10^{-2}$ & $10^{-4}$--$10^{2}$ & $10^{47}$--$10^{49}$ & $10^{48}$--$10^{52}$ & $10^{-2}$--$10^0$ \\
130603B & 8 & $0.1$--$1$ & $10^{-5}$--$10^0$ & $10^{-4}$--$10^1$ & $10^{47}$--$10^{49}$ & $10^{48}$--$10^{51}$ & -- \\
\hline
\end{tabular}
\caption{The range of physical parameters and energy factors found in the models that successfully matched the data (including radio observations). No value for $\kappa$ is shown for GRB 051221A and GRB 130603B because these bursts do not contain EE. The reduced $\chi^2$ thresholds are also shown. \label{tab:results}}
\end{center}
\end{table*}

\section{Modelling results}\label{sec:results}

The order of magnitude parameter search returned a variety of viable combinations across the four GRBs. Each was inspected by eye to ensure that no upper limits were violated and that the model was consistent with (i.e. fainter than) the early X-ray emission, since neither of these things were factored into the $\chi^2$ value. The fit each model gave to other optical and UV observations was also inspected for consistency, and those that violated upper limits or provided a poor match to the data were rejected. 16 models were found for GRB 051221A, 6 models were found for GRB 060614, 21 models were found for GRB 070714B, and 17 models were found for GRB 130603B. Example fits for each GRB are shown in Fig.~\ref{fig:example}. The X-ray rebrightening at around $10^{-2}$~d in GRB 070714B is interpreted as an X-ray flare \citep[e.g.][]{Gompertz13} due to its short time-scale and apparent discrepancy with the R-band light curve. This is supported by a spectral hardening shown by the photon index fit on the UKSSDC burst analyser\footnote{www.swift.ac.uk/burst\_analyser} \citep{Evans07,Evans09}

For each parameter combination, the $\chi^2$ values are calculated separately for the X-ray and R-band light curves. These are plotted against each other, and we make $\chi^2$ cuts at both frequencies that return a sample of the best fits for each GRB. This method prevents the much better sampled X-ray light curve from dominating the selection threshold, as would be the case for a combined reduced $\chi^2$ cutoff. The reduced $\chi^2$ limits that result from the combination of $\chi^2$ cutoffs for each burst are shown in Table~\ref{tab:results}. The large variations in these limits are a reflection on how constraining the available X-ray and R-band data are to the models; since the fitting procedure is a simple order of magnitude search rather than a least-squares fit, light curves with larger numbers of data points will be much less forgiving on the models applied. A finer parameter search would reduce $\chi^2$. The reduced $\chi^2$ limits for all four bursts could also be made more uniform with least-squares fitting; however, this approach leads to very large parameter uncertainties, as previously discussed.

\begin{figure*}
\begin{center}
\includegraphics[width=7.6cm]{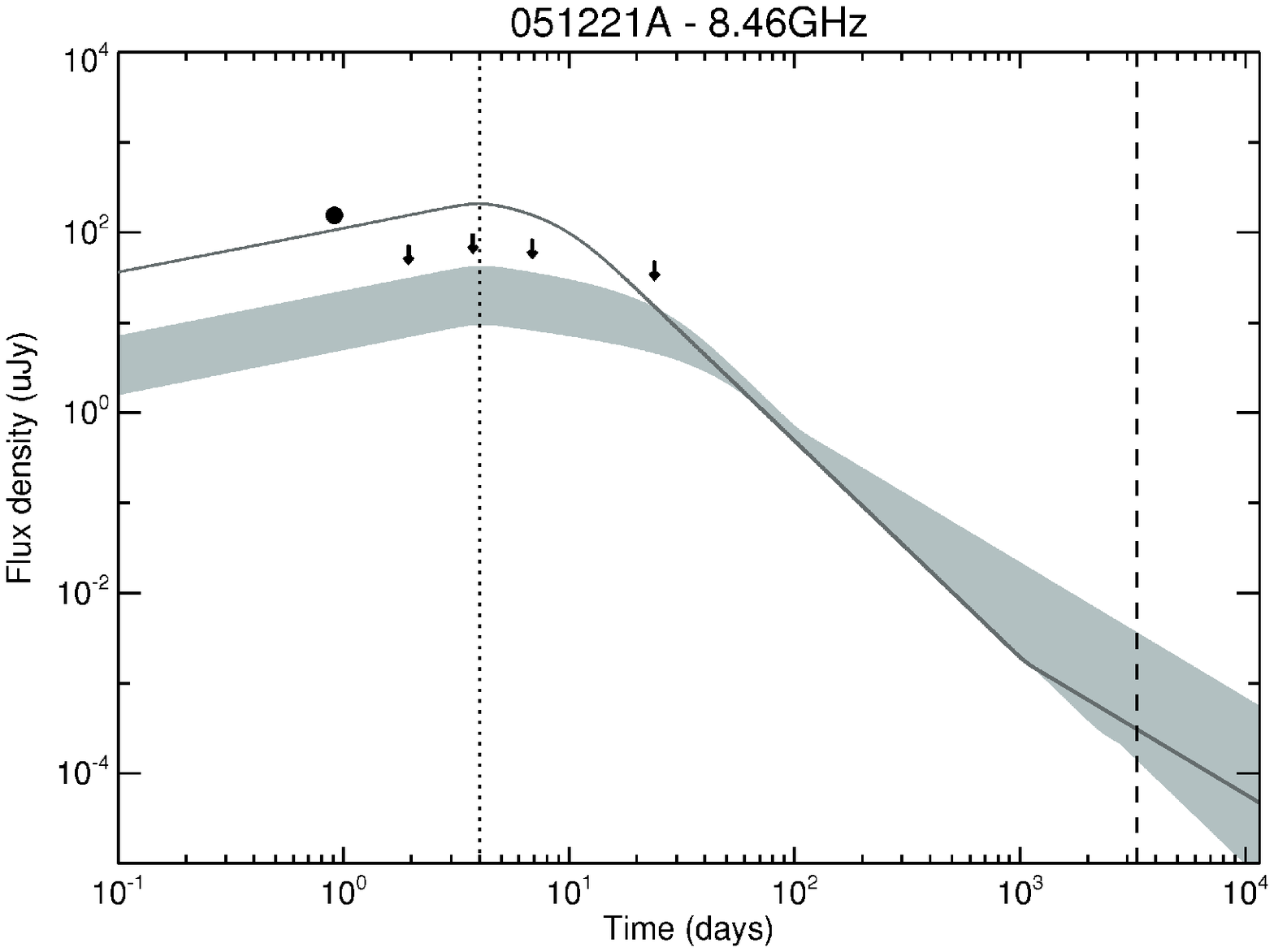}
\includegraphics[width=7.6cm]{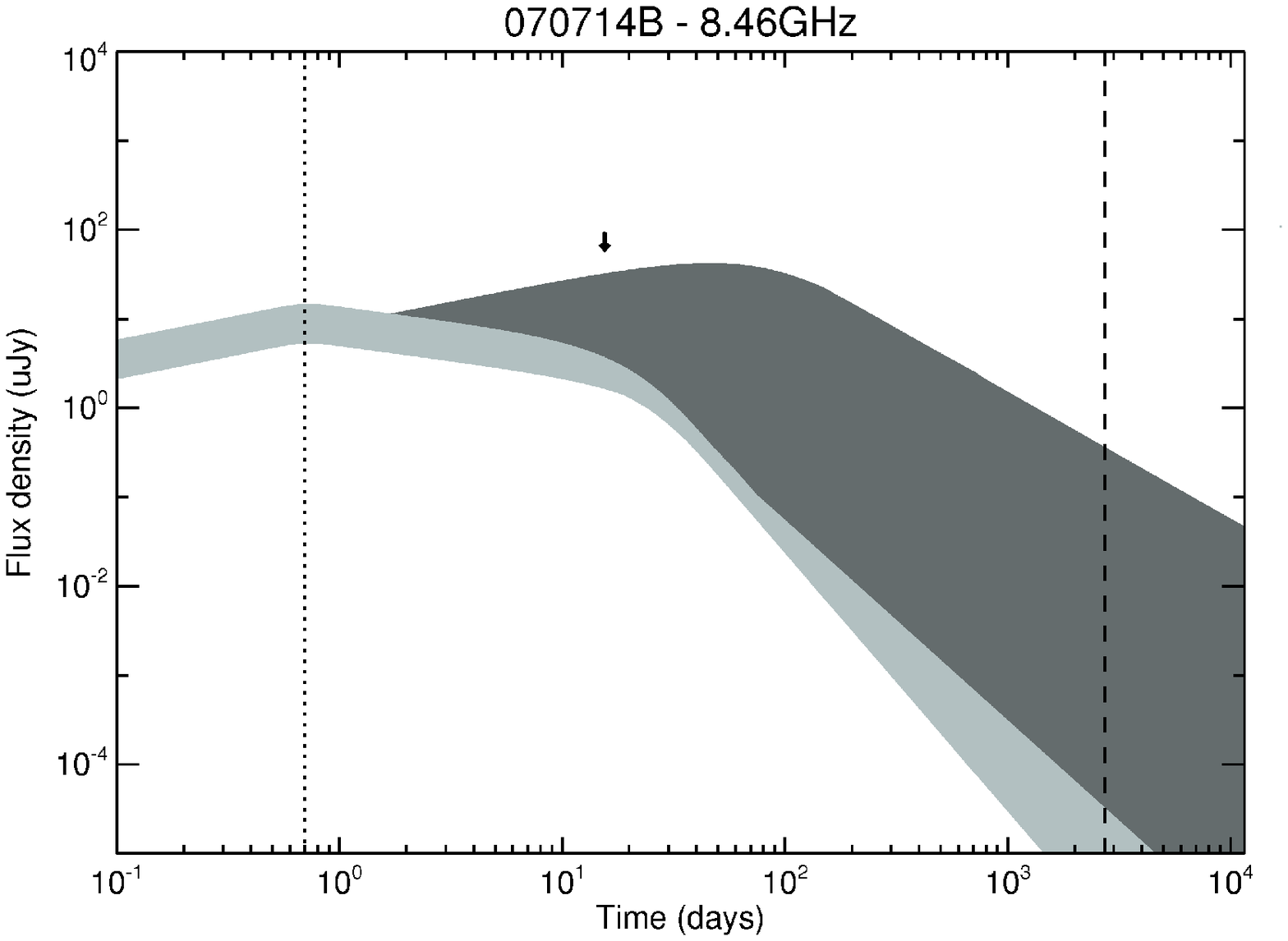}
\includegraphics[width=7.6cm]{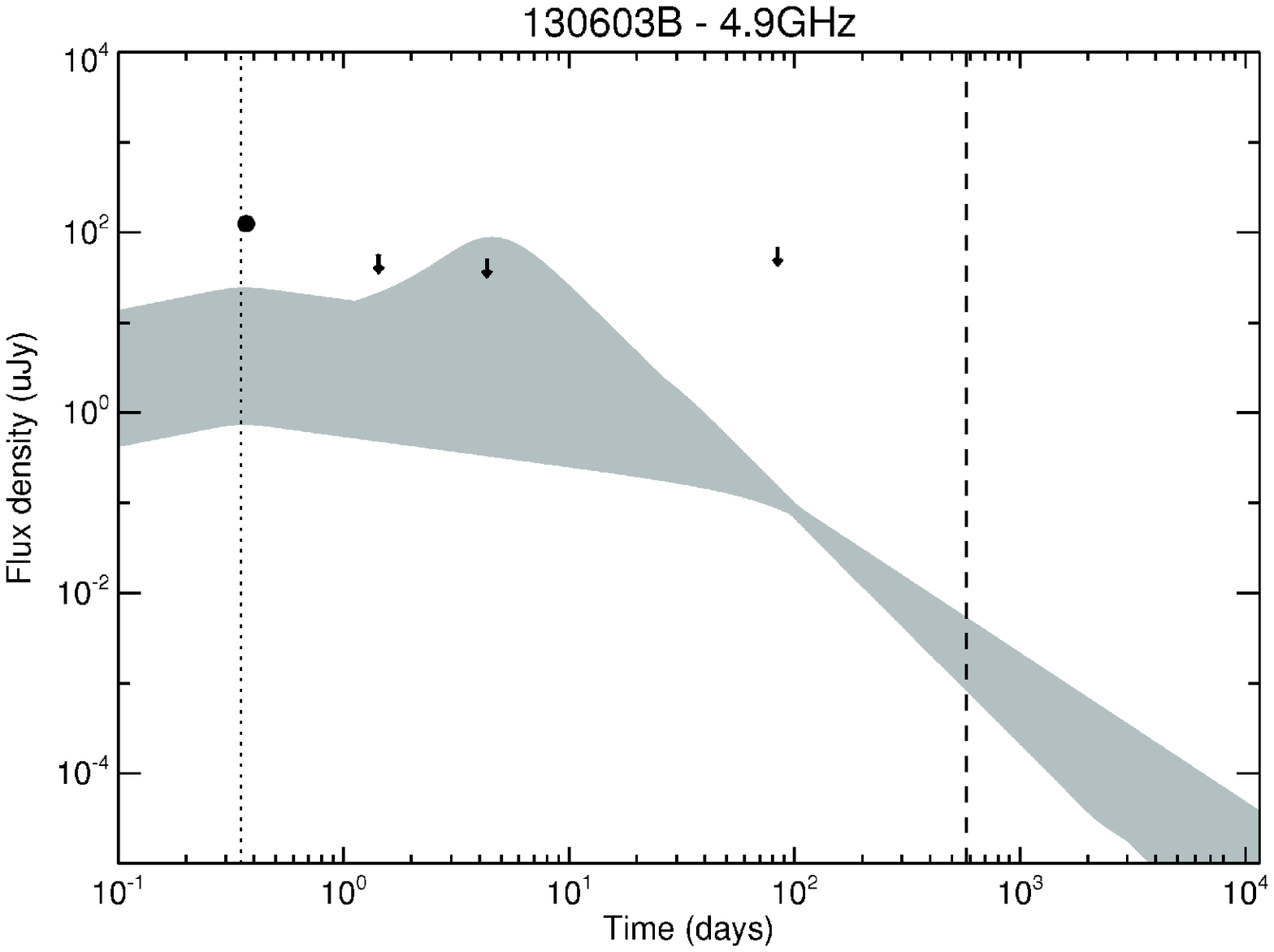}
\includegraphics[width=7.6cm]{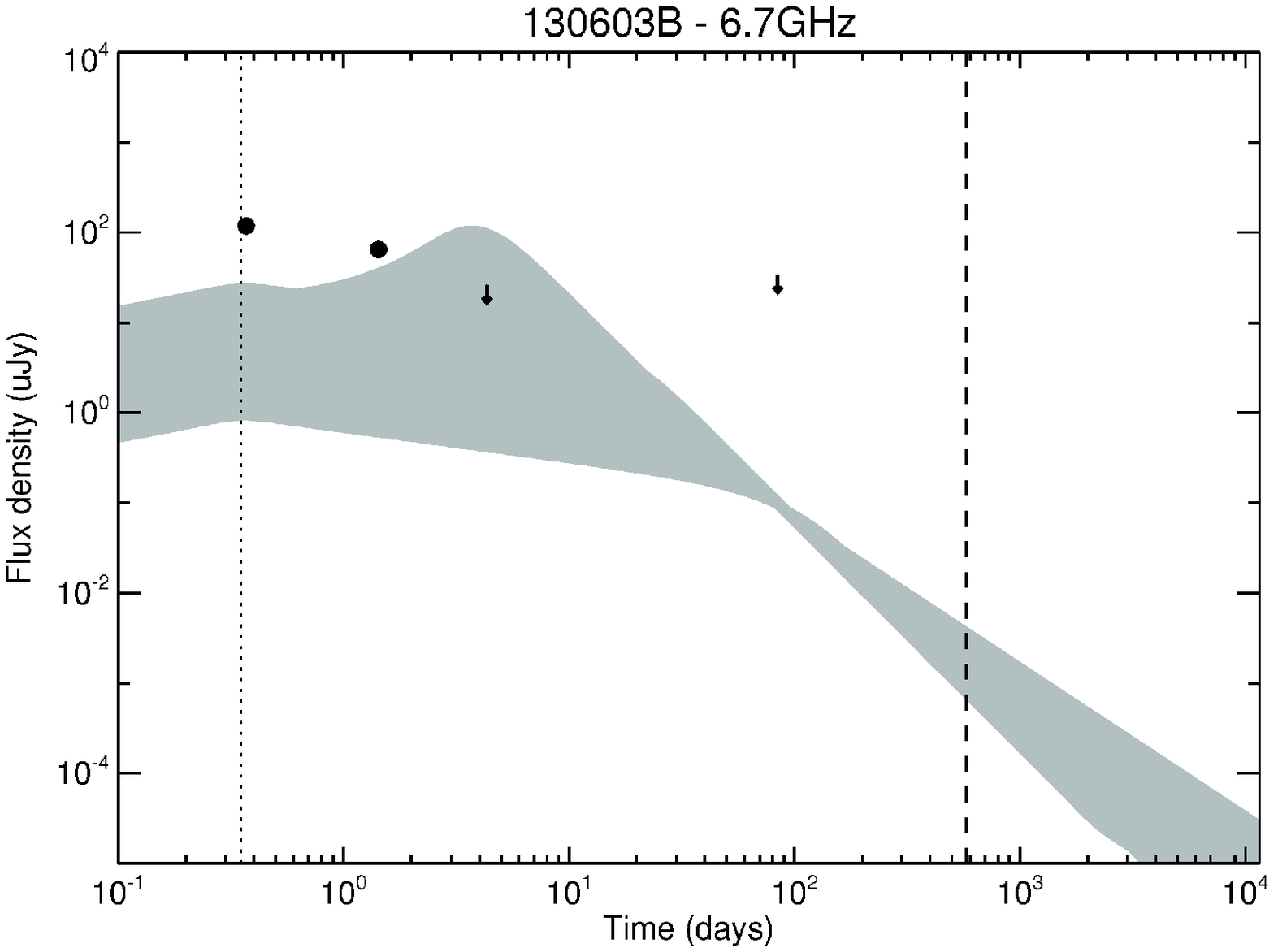}
\includegraphics[width=7.6cm]{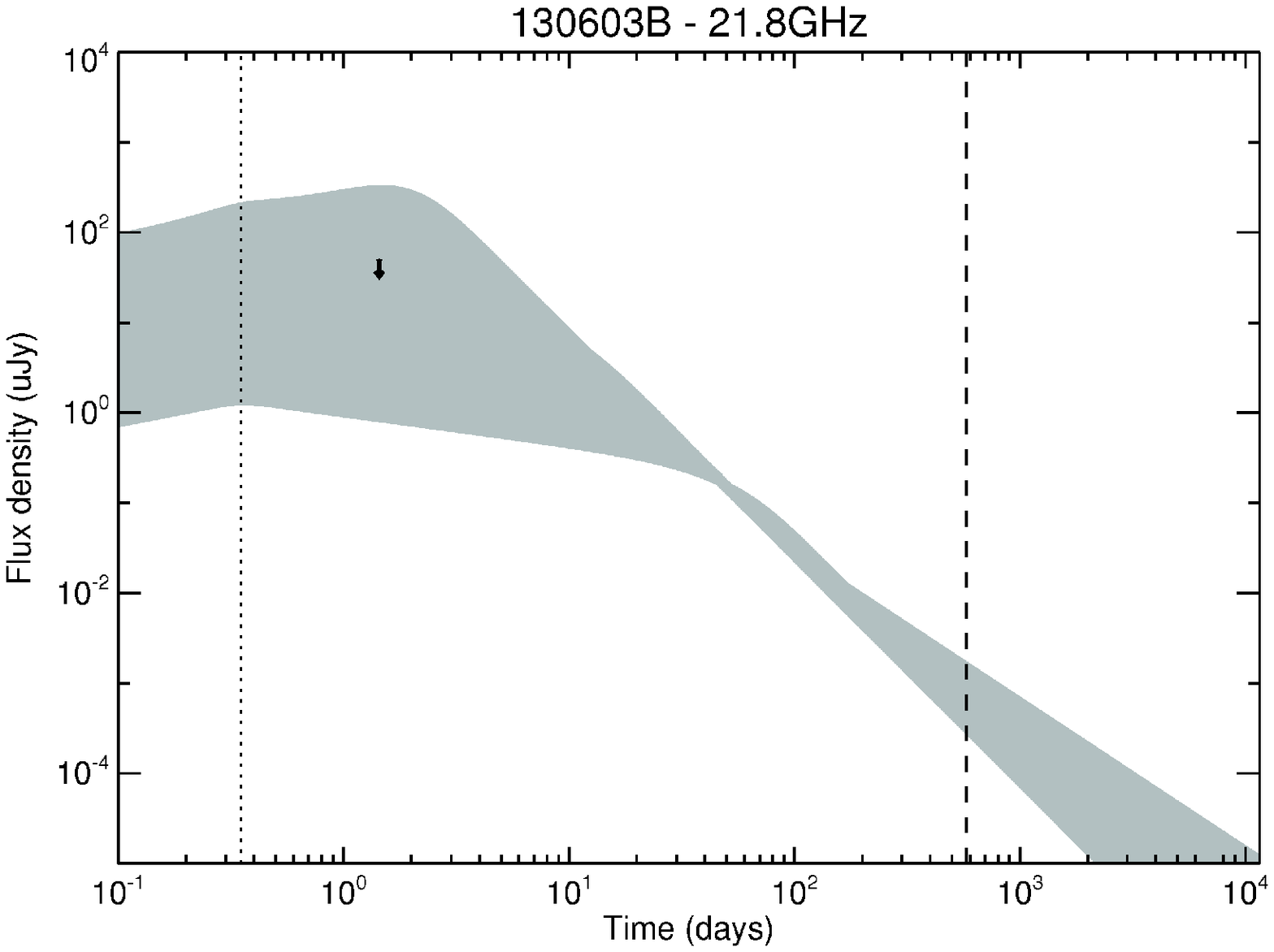}
\end{center}
\caption{The radio detections and upper limits available for our sample of four GRBs. The light grey region shows the range of fluxes described by the light curves of the model fits that are consistent with the broadband data. See Section~\ref{sec:discussion} for a discussion in the apparent upper limit violations in GRB 130603B. The dark grey line in GRB 051221A shows the model that comes closest to matching the observations at $8.46$~GHz, and its inconsistency with the upper limits illustrates the probable need for reverse shock emission at early times. The dark grey region in GRB 070714B shows the additional range of predictions resulting from the models with no jet break that are consistent with observations. The vertical black dotted line shows the position of the required jet break for GRB 051221A and GRB 130603B, and the position of the earliest jet break allowed by the data for GRB 070714B. The vertical black dashed line marks the 1$^{\rm st}$ of January 2015 for reference. \label{fig:radio}}
\end{figure*}

Three of the four GRBs (051221A, 070714B and 130603B) also feature radio detections and upper limits. The presence of radio observations helps to narrow the parameter space, with varying degrees of severity depending on how constraining the observation is. They are shown in Fig.~\ref{fig:radio}, plotted alongside the region described by the various light curves from the surviving models, shown in grey. GRB 051221A is the most constrained by radio observations; eight more models were ruled out due to the consecutive upper limits at $8.46$~GHz, including all models not featuring a jet break. The dark grey line shows the model that best matches the radio detection for this burst, but it is at odds with the upper limits. In GRB 130603B, the most luminous models appear to violate some of the upper limits; however, these can be retained due to the possible influence of radio scintillation \citep{Frail97,Goodman97} which can explain discrepancies in isolated cases. Forward shock emission appears to have some difficulty in matching the radio detections in this burst and GRB 051221A, and possible reasons for this are discussed in Section~\ref{sec:discussion}. In GRB 070714B, the upper limit is not at all constraining to the physical parameter space.

For each GRB, a fairly wide range of parameters was found. It is immediately obvious from Table~\ref{tab:results} that a high value of $\epsilon_{\rm e}$ is required in all cases, otherwise the model emission is too faint to match what is observed in both X-rays and optical bands. It should be noted that while we give broad ranges for the physical parameter values, these values only work to reproduce the data in specific combinations. Two of the four GRBs (060614 and 070714B) have values for $p$ that are consistent within the $90$ per cent confidence interval with the late-time photon index $\Gamma$ from the UKSSDC spectrum repository (Table~\ref{tab:datatable}). The other two lie between the values gained when using the $\nu < \nu_{\rm c}$ and $\nu_{\rm c} < \nu$ closure relations, indicating some evidence for a cooling break. Although there may be models with and without cooling breaks for each burst when using different parameter combinations, the example best fits in Fig.~\ref{fig:example} support this statement, since GRB 051221A and GRB 130603B exhibit late X-ray cooling breaks, while the other two GRBs do not. In two bursts (060614 and 070714B), the data appear to show the peak frequency passing through the R-band, with the cooling break sitting well above the X-ray band. This is a feature of all GRB 060614 models, and is very constraining to the physical parameters because it breaks the $\nu_{\rm m}$ -- $F_{\nu,\rm max}$ degeneracy.

\section{Modelling discussion}\label{sec:discussion}

In some cases, most notably GRB 130603B, we have retained models that appear to violate radio upper limits. This is based on radio scintillation arguments \citep{Frail97,Goodman97}, where isolated detections and upper limits could be located at scintillation maxima/minima, and therefore have larger errors than quoted. For this reason, the handful of models that do not obey the upper limit in GRB 130603B have been left in the predictions. In cases like the $8.46$~GHz light curve for GRB 051221A, however, consecutive limits are unlikely to have all occurred at scintillation minima, and so the models that passed above more than one of them (including all the no jet break models) were rejected.

This causes a problem when trying to explain the earlier radio detection with forward shock emission alone; the only model that comes close in GRB 051221A, shown by the dark grey line in Fig.~\ref{fig:radio}, is inconsistent with three of the four radio upper limits. While it could be argued that the single detection in GRB 051221A is itself due to scintillation, the situation is even worse in GRB 130603B, where consecutive detections at $6.7$~GHz and a further observation at $4.9$~GHz cannot be matched by models without rising above multiple upper limits. The natural explanation for this is the presence of a reverse shock propagating backwards through the ejected material. A reverse shock could produce a radio flare, providing a match to the data while still being masked beneath the forward shock emission at higher frequencies \citep{Kulkarni99,Sari99b,Nakar05,vanEerten14}. Reverse shocks are believed to have been observed in both LGRBs \citep[e.g.][]{Akerlof99,Chandra10,Anderson14} and SGRBs (e.g. \citealt{Soderberg06} for GRB 051221A).

Our model includes only the most basic features of the magnetar central engine; we have not incorporated reverse shock emission or other sources of radiation \citep[e.g. kilonova emission,][]{Piran13,Tanvir13}. Our aim was to show that a physically motivated, self-consistent central engine, in which a newly formed magnetar injects energy into an expanding forward shock as it loses angular momentum, can be reconciled with the longer wavelength (optical, IR, radio) observations of SGRBs, as well as just the X-ray light curves as is usually done. For this reason, and given the roughness of our fitting routine, the fact that the light curves at all frequencies are well recreated by this bare-bones model is encouraging.

\citet{Soderberg06} modelled the afterglow of GRB 051221A. Our results are in agreement with theirs, except that we find a much wider range in $\epsilon_B$ ($10^{-4}$--$10^{-1}$ in this work, compared to $0.12$--$1/3$ in \citealt{Soderberg06}) and $n_0$ ($10^{-4}$--$10^1$~cm$^{-3}$ in this work, compared to $(0.5$--$2.4) \times 10^{-3}$~cm$^{-3}$ in \citealt{Soderberg06}). This narrow range is likely due to the inclusion of a reverse shock in their modelling, and indeed their forward shock only parameter ranges are much broader, although still narrower than what we find. GRB 051221A was also modelled by \citet{Burrows06}, who obtain a low- and high-density fit, giving a range of $10^{-4}$~cm$^{-3} \leq n_0 \leq 0.1$~cm$^{-3}$ which is in agreement with our findings, and similar to the forward shock only results of \citet{Soderberg06}. Both studies find narrow jets, consistent with our range (Section~\ref{sec:energy}), and jet break times of $4$--$5$~d. \citet{Fan06} also fitted the magnetar model to the broadband observations of GRB 051221A, finding a family of physical parameters within our range.

No broadband modelling has been done on GRB 070714B, but \citet{Xu09} fitted a model featuring power-law energy injection to GRB 060614, and found a fit with $\epsilon_{\rm e} \sim 0.12$, $\epsilon_B \sim 2\times10^{-4}$, and $n_0 = 0.04$~cm$^{-3}$, in agreement with our range of parameters. By fitting power-law models to the R-band light curves \citep{DellaValle06} and a combination of X-ray and optical bands \citep{Mangano07}, two previous studies have found a jet break at $\sim 1.3$~d in GRB 060614, consistent with what we find at $1.1$~d.

The broadband afterglow of GRB 130603B was modelled by \citet{Fong14}. As in 051221A, our derived range of density values is wider, extending two orders of magnitude lower than \citet{Fong14}. Our $\epsilon_B$ range also extends down an order of magnitude further. These ranges highlight the large degeneracies in the parameters; $\epsilon_{\rm e}$ is confined to a relatively small range ($\sim$ one order of magnitude) because $\nu_{\rm m}$ and $F_{\nu,\rm max}$ are well constrained by the data, whereas $\nu_{\rm c}$ and $\nu_{\rm a}$ are often unconstrained, leading to a variety of acceptable parameter combinations. \citet{Fong14} find a jet break at $\approx 0.47$~d, and a jet opening angle in the range $4^{\circ}$ -- $14^{\circ}$, both of which are consistent with our own findings. Finally, the magnetar spin period and dipole field values calculated by \citet{Fong14} for the dipole spin-down injection case intersect with the line for GRB 130603B shown in Fig.~\ref{fig:magnetar}. \citet{Fan13} also showed that the magnetar model was capable of reproducing the broadband emission observed in GRB 130603B for one combination of physical parameters that lies within our range.

\subsection{Energetics}\label{sec:energy}

The radiative efficiency of a GRB is defined as \citep[cf.][]{Zhang07}
\begin{equation}
\zeta = \frac{E_{\gamma,\rm iso}}{E_{\gamma,\rm iso}+E_{k,\rm iso}}
\end{equation}
and gives a direct measure of how efficiently the total energy is converted into EM radiation. In our work, we calculate $E_k$ for each burst, which is the energy delivered to the afterglow emission site by the prompt impulse, and makes no assumption on geometry. The lower limit of $E_k$ is not at all constraining; the fit to the plateau emission depends much more on the luminosity of the dipole spin-down injection, $L$. Values for $E_k$ of $10^{48}$~erg and below are indistinguishable from one another, and for a given $E_{\gamma,\rm iso}$ will just represent an asymptotic approach to a radiative efficiency of $1$, which is unphysical. The upper limits of $E_k$ are far more important, since they are constrained by observation in that too much energy contribution will drive the model fluxes up above what is observed, and will mask the plateau feature in cases where $E_{\rm d}$ is negligible in comparison to $E_k$. The approximate maximum value of $E_{k,\rm iso}$ is given by assuming that the upper limit of $E_k$ came from a strongly beamed geometry with a beaming factor of $\sim 1000$, i.e. the upper limit of $E_{k,\rm iso}$ is as much as a thousand times greater than the upper limit of $E_k$. The radiative efficiency can then be used to calculate the implied opening angle \citep[cf.][]{Racusin09}:
\begin{equation}
\theta_j = 0.057t_{\rm jb}^{3/8}\bigg(\frac{3.5}{1+z}\bigg)^{3/8}\bigg(\frac{\zeta}{0.1}\bigg)^{1/8}\bigg(\frac{n_0}{E_{\gamma,\rm iso,53}}\bigg)^{1/8}.
\end{equation}
The range of calculated efficiencies and opening angles are shown in Table~\ref{tab:jets}. The derived efficiencies are consistent with \citet{Zhang07}, who found typical values of $< 10$ per cent in their sample. Note that for the two EE GRBs, these calculations may be affected by the energy contribution of EE.  We find that GRB 060614 tends to demand higher values of $\epsilon_{\rm e}$ and $\zeta$ than the other bursts, which is symptomatic of its more luminous and longer lasting afterglow plateau putting extra demands on the available energy. The derived opening angles are consistent with the results of \citet{Ryan15}. Their results (in degrees) are $26.0^{+1.80}_{-2.20}$ for GRB 051221A, $17.0^{+7.08}_{-4.93}$ for GRB 060614 and $19.1^{+6.38}_{-6.38}$ for GRB 070714B, where the errors are $1\sigma$. These values were obtained by fitting to the X-ray light curves only.

\begin{table}
\begin{center}
\begin{tabular}{lccc}
\hline
GRB & $E_{\gamma,\rm iso}$ & $\zeta$ & $\theta_j$ \\
& (erg) & & (deg) \\
\hline
051221A & $1.5 \times 10^{51,a}$ & $\ge 1.5 \times 10^{-3}$ & $2.37$--$22.7^*$ \\
060614 & $2.5 \times 10^{51,b}$ & $\ge 2.0 \times 10^{-2}$ & $1.62$--$19.7$ \\
070714B & $1.6 \times 10^{51,c}$ & $\ge 1.6 \times 10^{-4}$ & $0.87$--$14.4^*$ \\
130603B & $1.0 \times 10^{51,d}$ & $\ge 1.0 \times 10^{-3}$ & $1.01$--$10.0$ \\
\hline
\end{tabular}
\caption{Calculated minimum radiative efficiencies and ranges of opening angles. $^a$\citet{Soderberg06}; $^b$\citet{Mangano07}; $^c$\citet{Graham09}; $^d$\citet{Fong14}. $^*$Calculated from a minimum $t_{\rm jb}$; value rises with increasing $t_{\rm jb}$. \label{tab:jets}}
\end{center}
\end{table}

\subsection{Magnetar properties}

\begin{figure}
\begin{center}
\includegraphics[width=8.8cm]{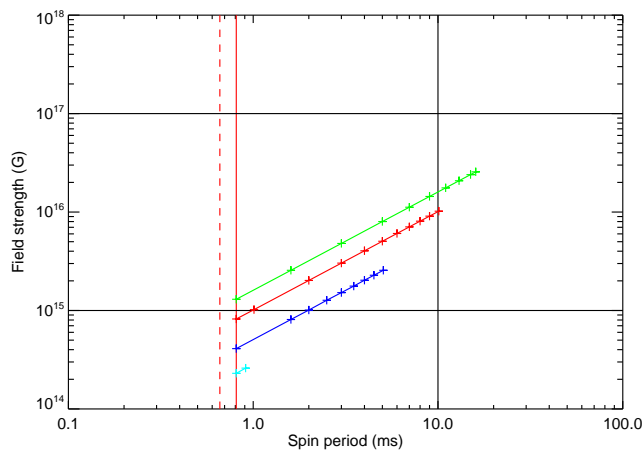}
\end{center}
\caption{Magnetar spin period and dipole field strength combinations that satisfy the luminosity limits and $T_{\rm em}$ values of the four GRBs. Blue -- GRB 051221A; light blue -- GRB 060614; red -- GRB 070714B; green -- GRB 130603B. The solid (dashed) vertical red line marks the spin break-up period for a $1.4$ ($2.1$)~M$_{\odot}$ NS \citep{Lattimer04}. The vertical black line represents the maximum allowed spin period at birth, based on the conservation of angular momentum of a white dwarf binary merger \citep{Usov92}. The lower horizontal limit marks the minimum magnetic field required to produce a GRB observable in the gamma band \citep{Thompson07} and the upper limit is the nominal threshold for fast field decay. \label{fig:magnetar}}
\end{figure}

The well constrained value of $T_{\rm em}$ and the results for $L$ mean that the magnetar properties can be approximated. These approximations assume $100$ per cent efficiency and isotropic emission, which is unlikely to be the case; however, in energetic terms a reduction in one compensates for a reduction in the other, and the large starting uncertainty associated with a simple order of magnitude search for $L$ means that this assumption is sufficiently accurate in the context of other sources of error. The range of magnetar spin periods and field strengths is illustrated in Fig.~\ref{fig:magnetar}. These properties are not well constrained in three out of four bursts due to the degeneracy created by the wide range of physical parameters, as well as uncertainties in measuring the dipole plateau due to contamination from the prompt and EE components. The normalization of the $B$--$P$ relation is set by the value of $T_{\rm em}$ for each burst, and the suitable combinations run from the minimum spin break-up period up to the point at which the plateau becomes too faint for a good fit, at around $L = 10^{47}$~erg~s$^{-1}$.

The implications for EE in GRB 060614 and GRB 070714B are not well defined. While a range of energies that work in the context of the light curves can be found, the physical interpretation is not constrained in terms of beaming or efficiency, save that the results lie in the region found here. One central engine capable of providing such a result is a magnetic propeller \citep{Gompertz14}. The EE profile used here borrowed the luminosity curve for a $40$ per cent efficient isotropic propeller (without the prefactor $\kappa$), so the EE contribution is energetically consistent with the propeller model, however in this context it was used as a simple indicator of luminosity, and the restrictions it imposes on $P$ and $B$ of the underlying magnetar were not applied.

\begin{table}
\begin{center}
\begin{tabular}{lcc}
\hline
Telescope & Sensitivity & Reference \\
& ($\mu$Jy) & \\
\hline
\vspace{0.15cm}\emph{60~MHz:}\\
LWA1 & $38000$ & \citet{Ellingson13} \\
LOFAR & $5000$ & \citet{vanHaarlem13} \\
\hline
\vspace{0.15cm}\emph{150~MHz:}\\
MWA & $1200$ & \citet{Tingay13} \\
LOFAR & $300$ & \citet{vanHaarlem13} \\
\hline
\vspace{0.15cm}\emph{1.4~GHz:}\\
GMRT & $150$ & \citet{Ghirlanda14} \\
WSRT/Apertif & $50$ & \citet{Ghirlanda14} \\
ASKAP & $50$ & \citet{Ghirlanda14} \\
MeerKAT phase 1 & $9$ & \citet{Ghirlanda14} \\
MeerKAT phase 2 & $6$ & \citet{Ghirlanda14} \\
SKA phase 1 & $1$ & \citet{Ghirlanda14} \\
SKA phase 2 & $0.15$ & \citet{Ghirlanda14} \\
\hline
\vspace{0.15cm}\emph{15~GHz:}\\
AMI & $70$ & \citet{Zwart08} \\
VLA & $5$ & \citet{Ghirlanda14} \\
\hline
\vspace{0.15cm}\emph{100~GHz:}\\
CARMA & $900$ & \citet{Bock06} \\
ALMA & $6$ & [A] \\
\hline
\end{tabular}
\caption{Detection sensitivities for different instruments at the frequencies for which we calculate flux density prediction light curves. Limits are $5\sigma$ and assume a $12$~h integration time. [A] -- almascience.eso.org/proposing/sensitivity-calculator \label{tab:limits}}
\end{center}
\end{table}

\section{Implications for radio emission}\label{sec:predictions}

\begin{figure*}
\begin{center}
\includegraphics[width=7.6cm]{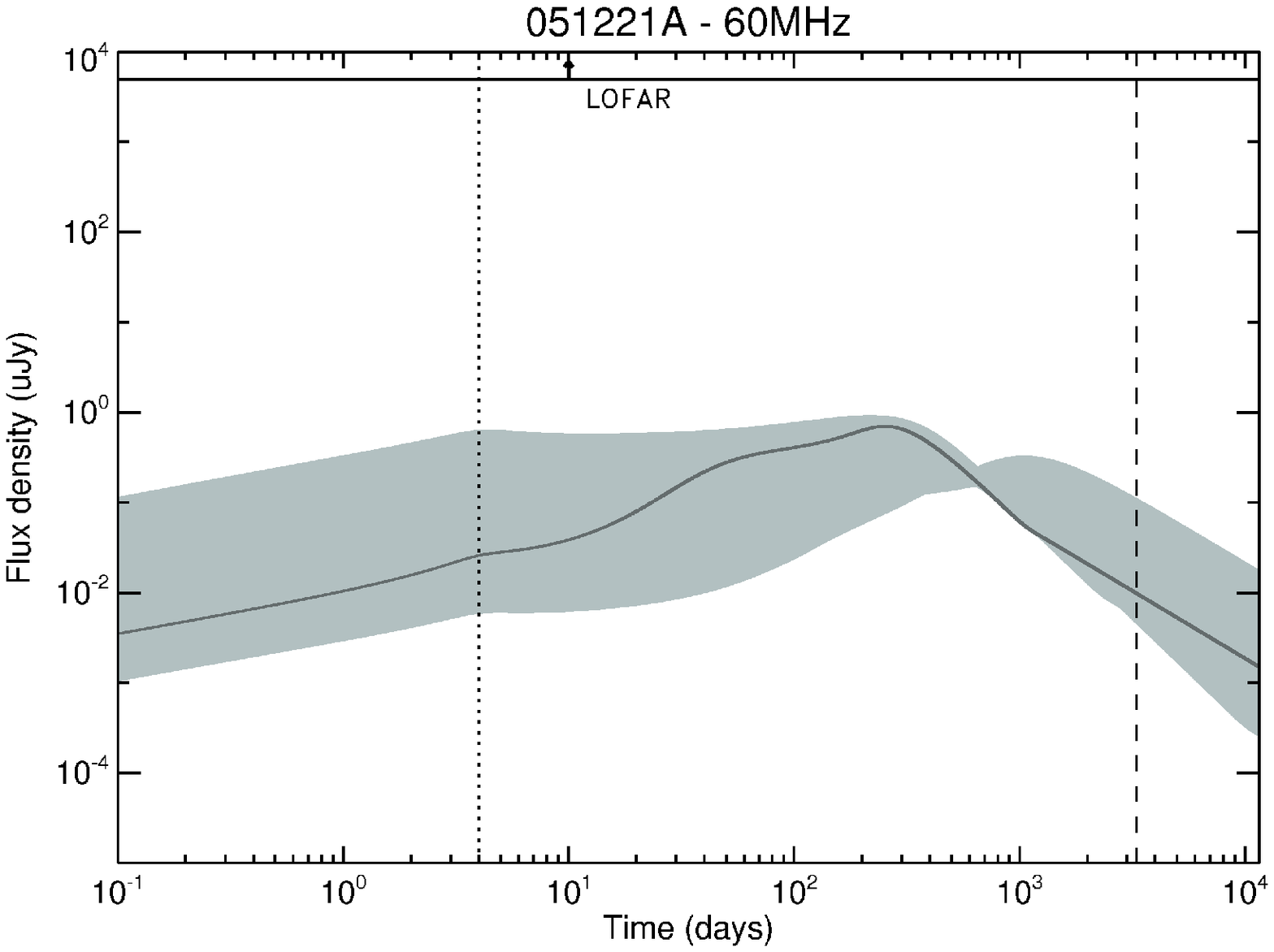}
\includegraphics[width=7.6cm]{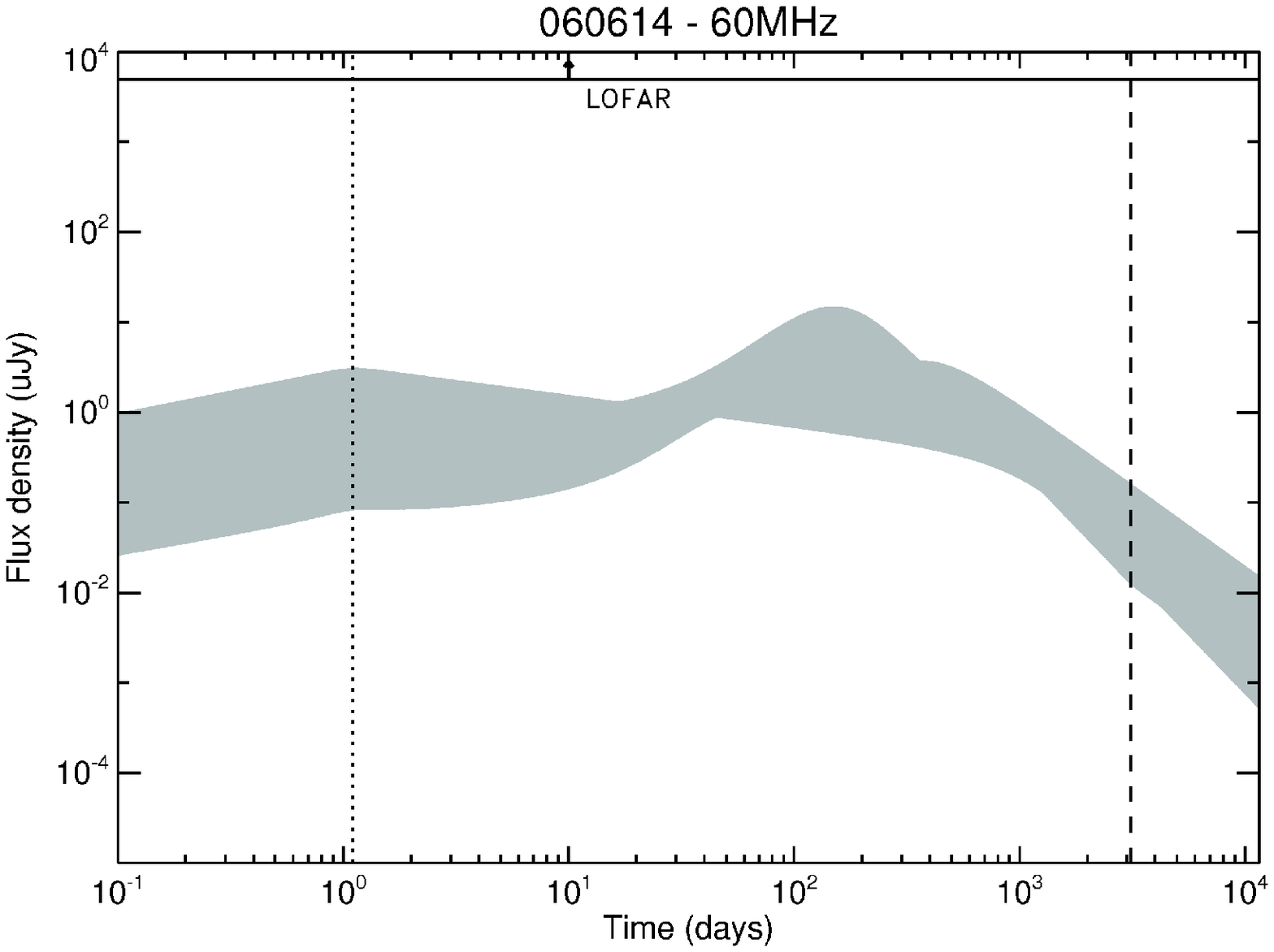}
\includegraphics[width=7.6cm]{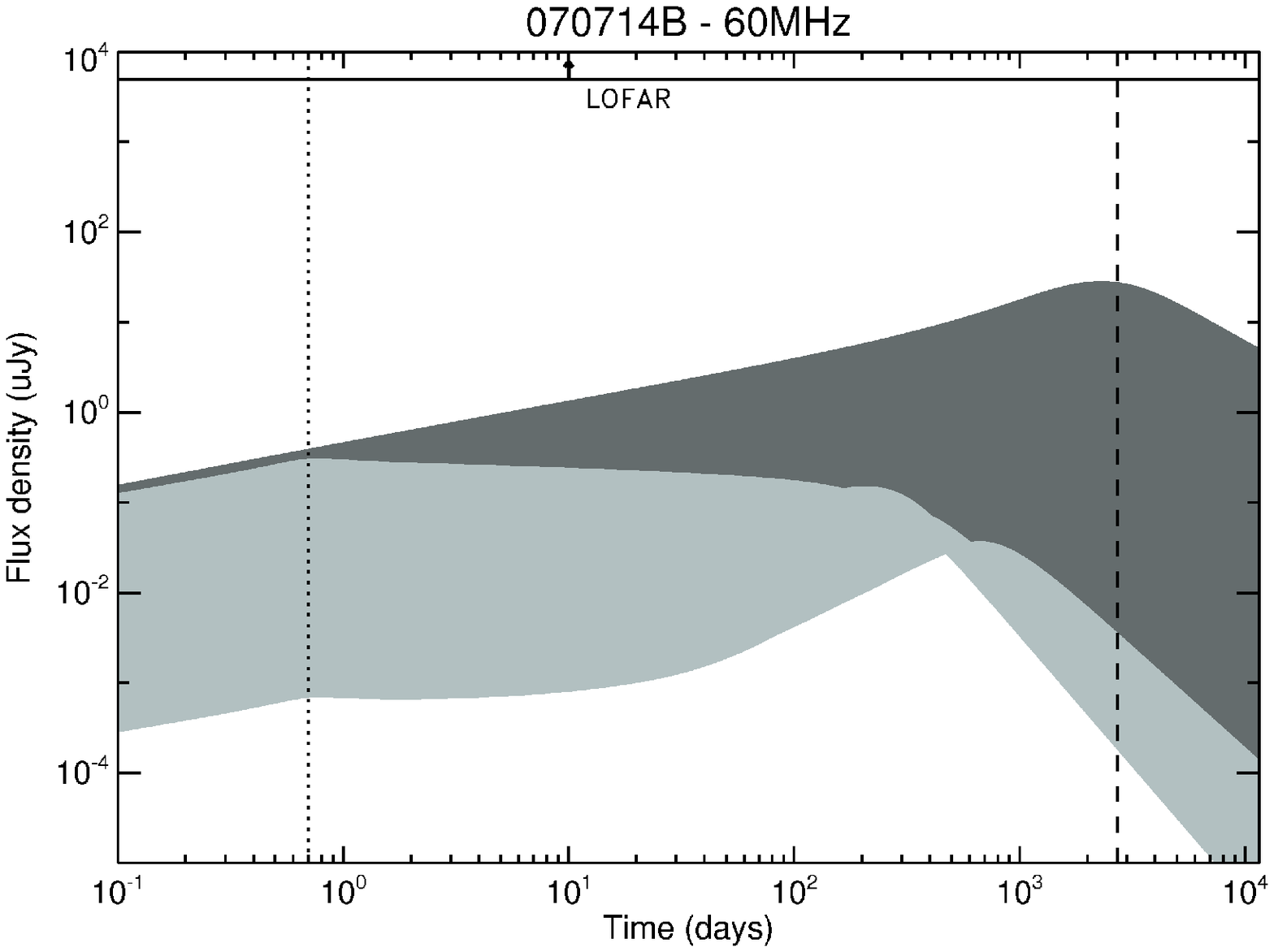}
\includegraphics[width=7.6cm]{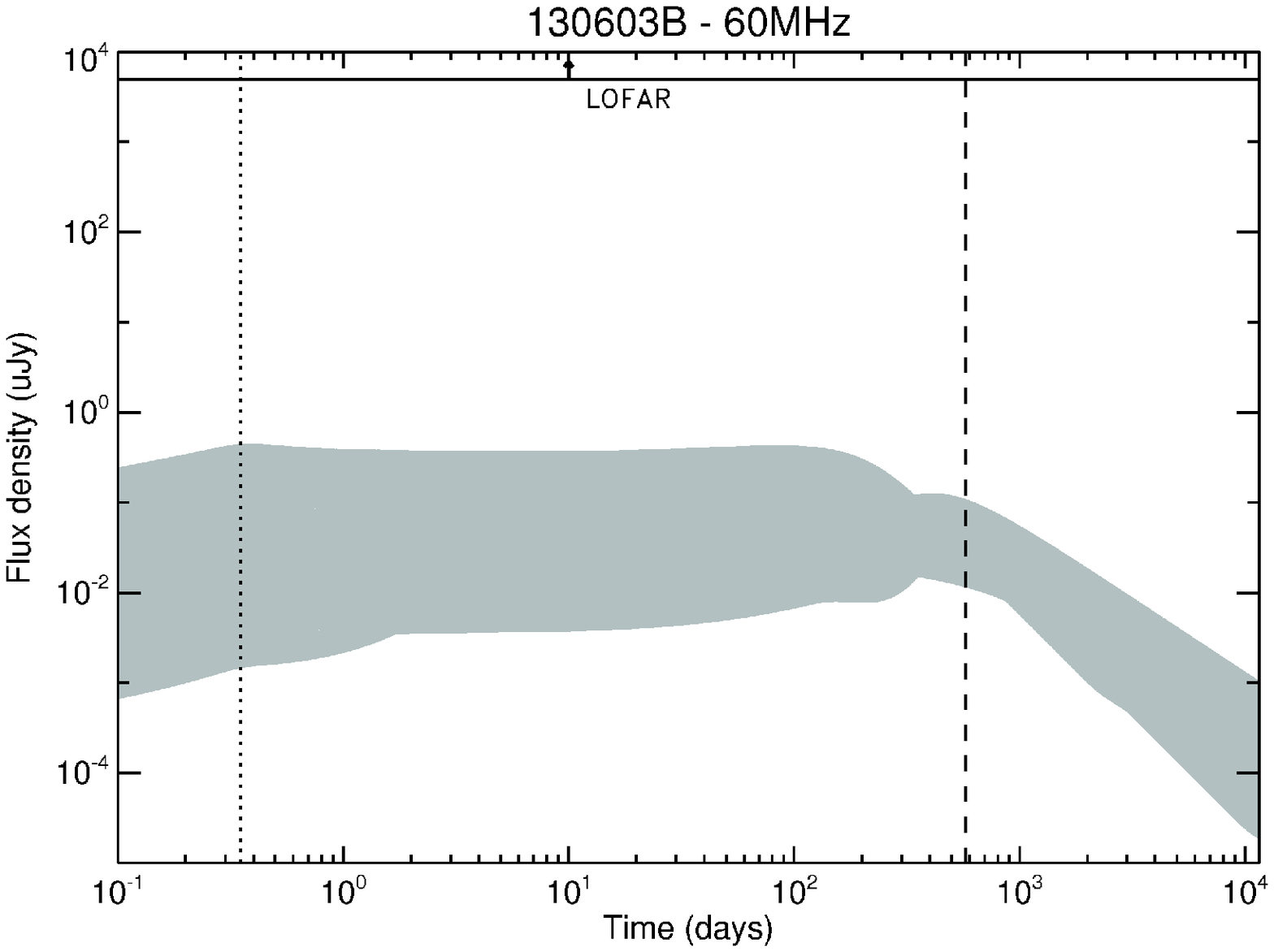}
\includegraphics[width=7.6cm]{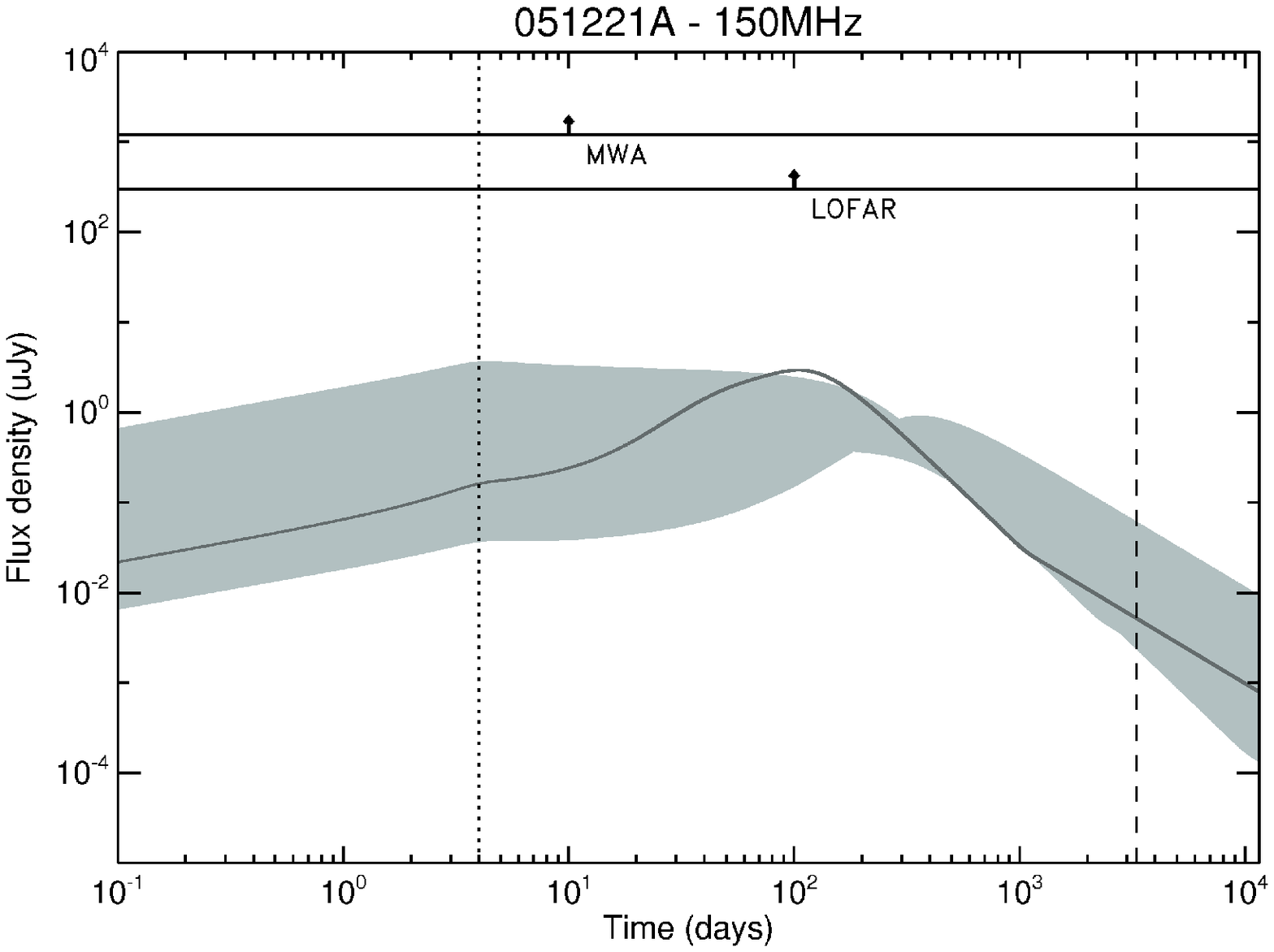}
\includegraphics[width=7.6cm]{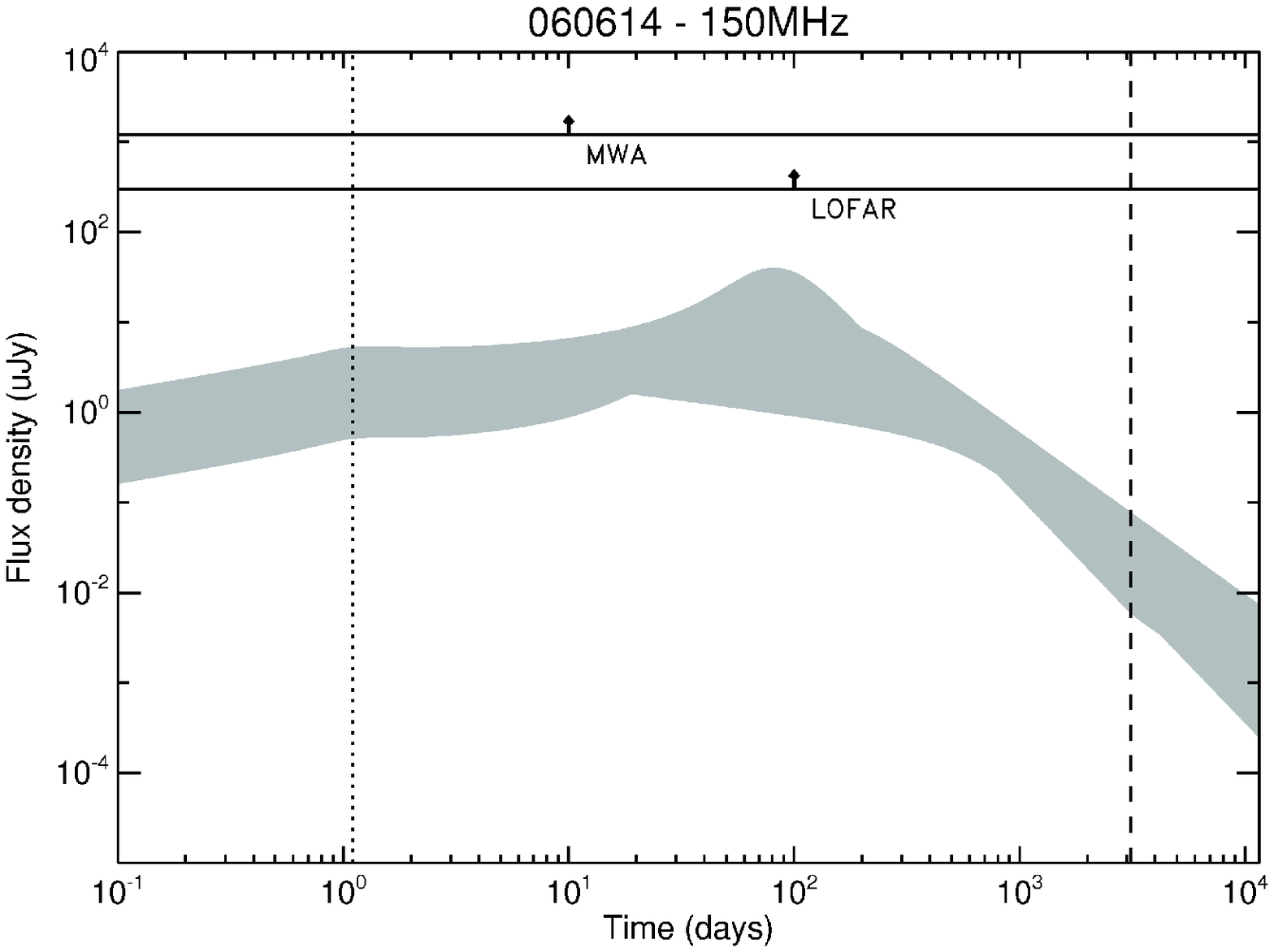}
\includegraphics[width=7.6cm]{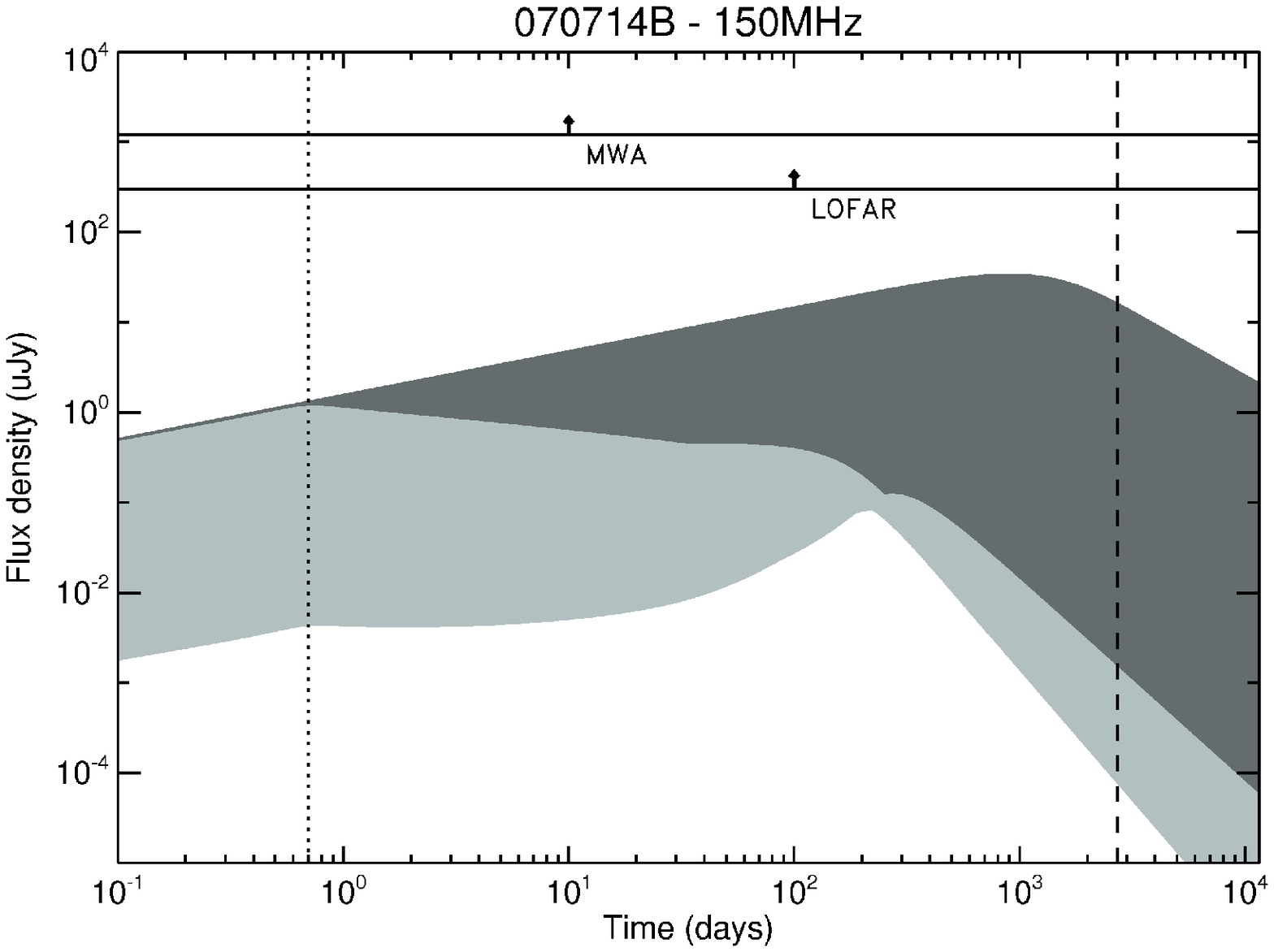}
\includegraphics[width=7.6cm]{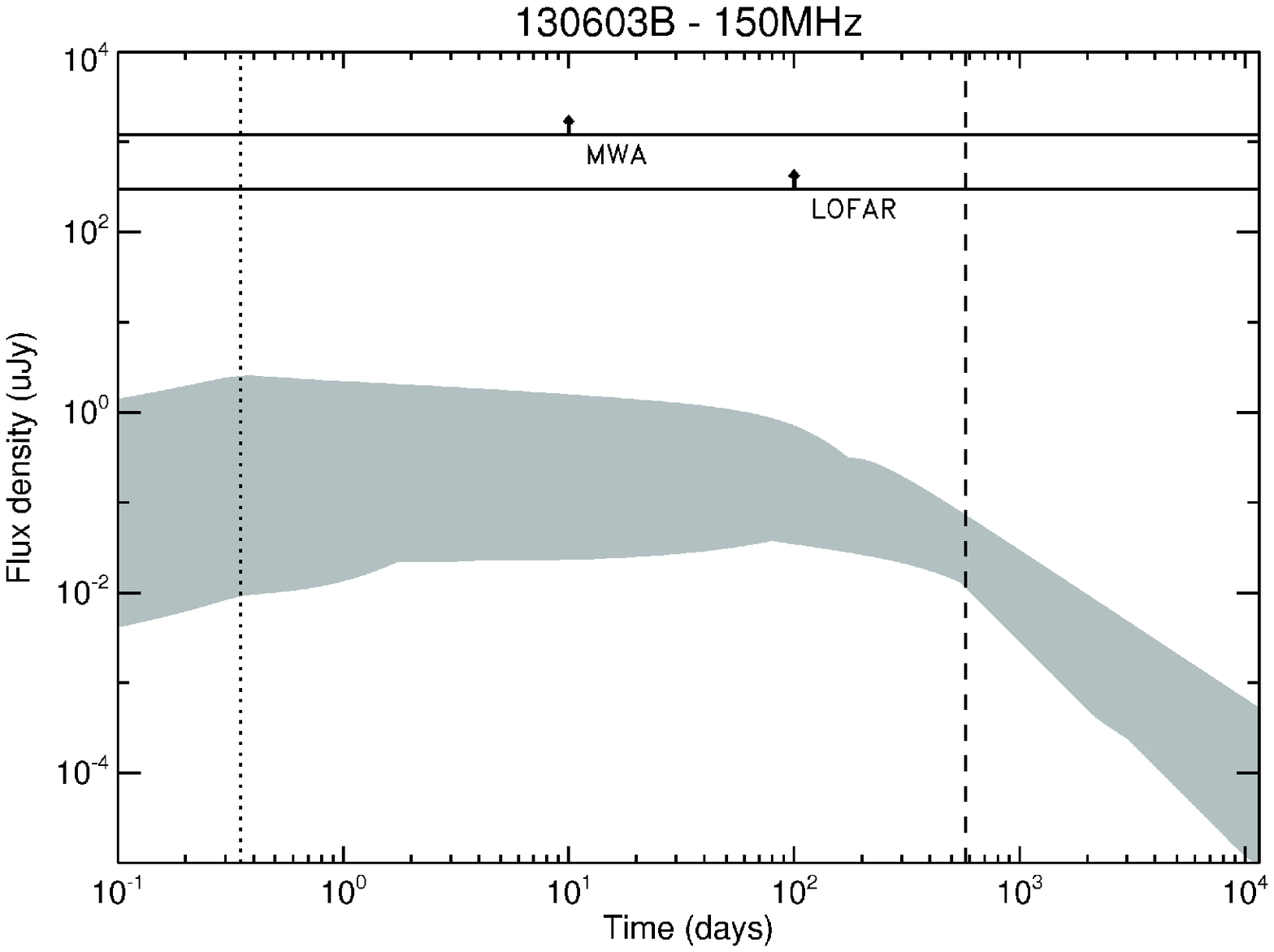}
\end{center}
\caption{Predicted flux density light curves at $60$ and $150$~MHz for the four GRBs in our sample. The dark grey line in GRB 051221A shows the model that comes closest to matching the observations at $8.46$~GHz in Figure~\ref{fig:radio}. The dark grey region in GRB 070714B shows the additional range of predictions resulting from the models with no jet break that are consistent with observations. The vertical black dotted line shows the position of the required jet break for GRB 051221A, GRB 060614 and GRB 130603B, and the position of the earliest jet break allowed by the data for GRB 070714B. The vertical black dashed line marks the 1$^{\rm st}$ of January, 2015, for reference. Selected limits from Table~\ref{tab:limits} are overplotted. \label{fig:results}}
\end{figure*}
\addtocounter{figure}{-1}
\begin{figure*}
\begin{center}
\includegraphics[width=7.6cm]{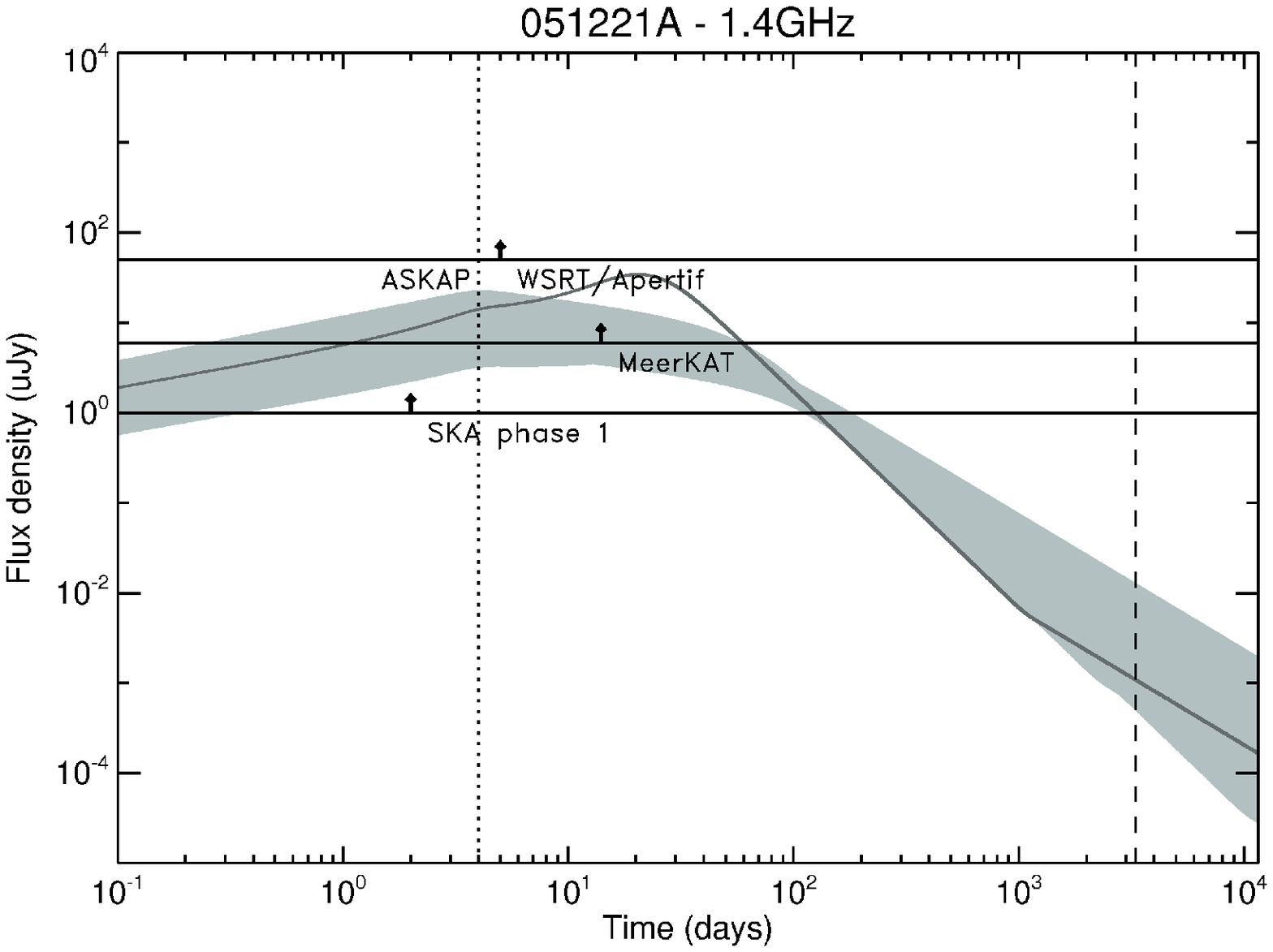}
\includegraphics[width=7.6cm]{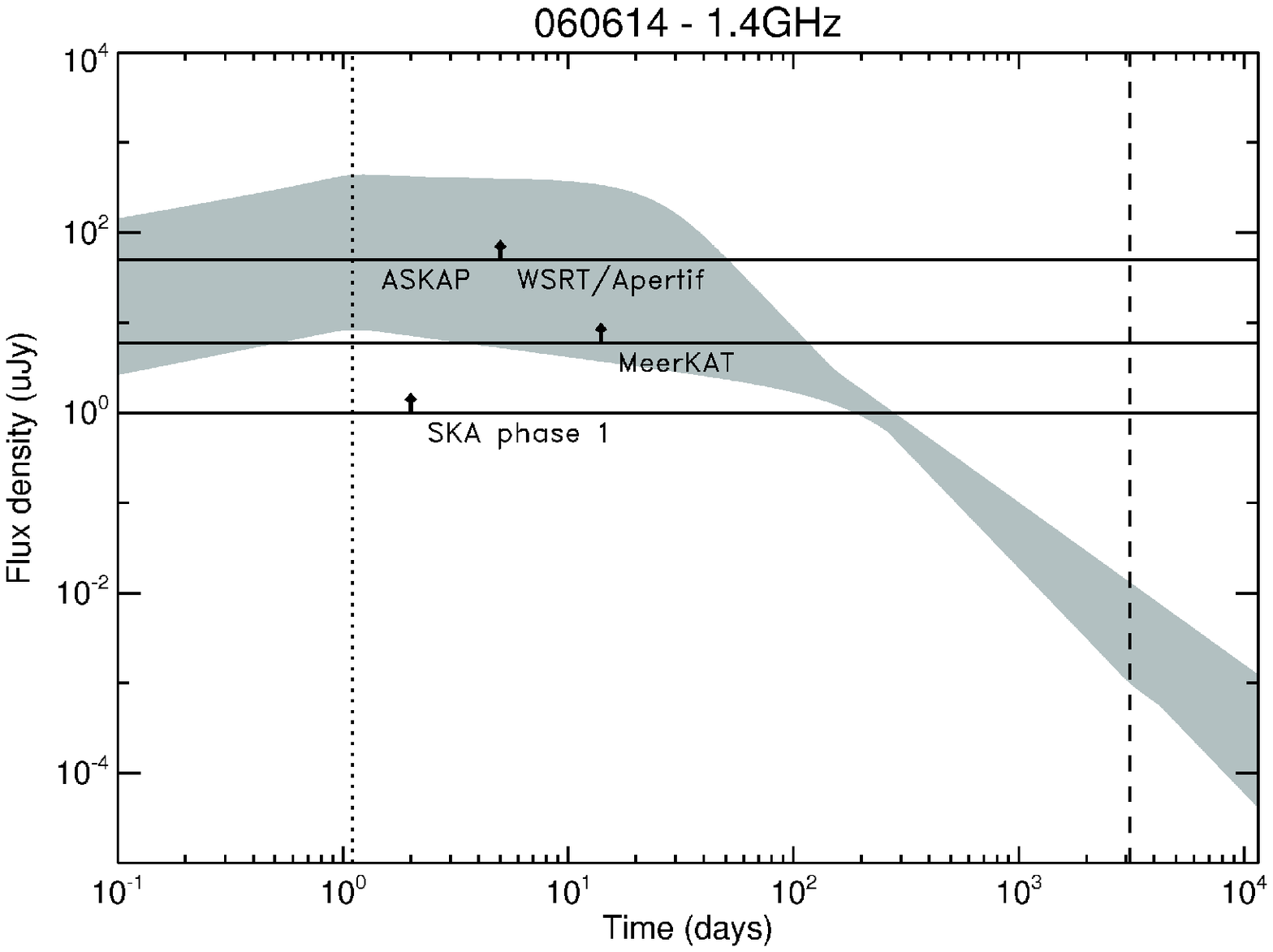}
\includegraphics[width=7.6cm]{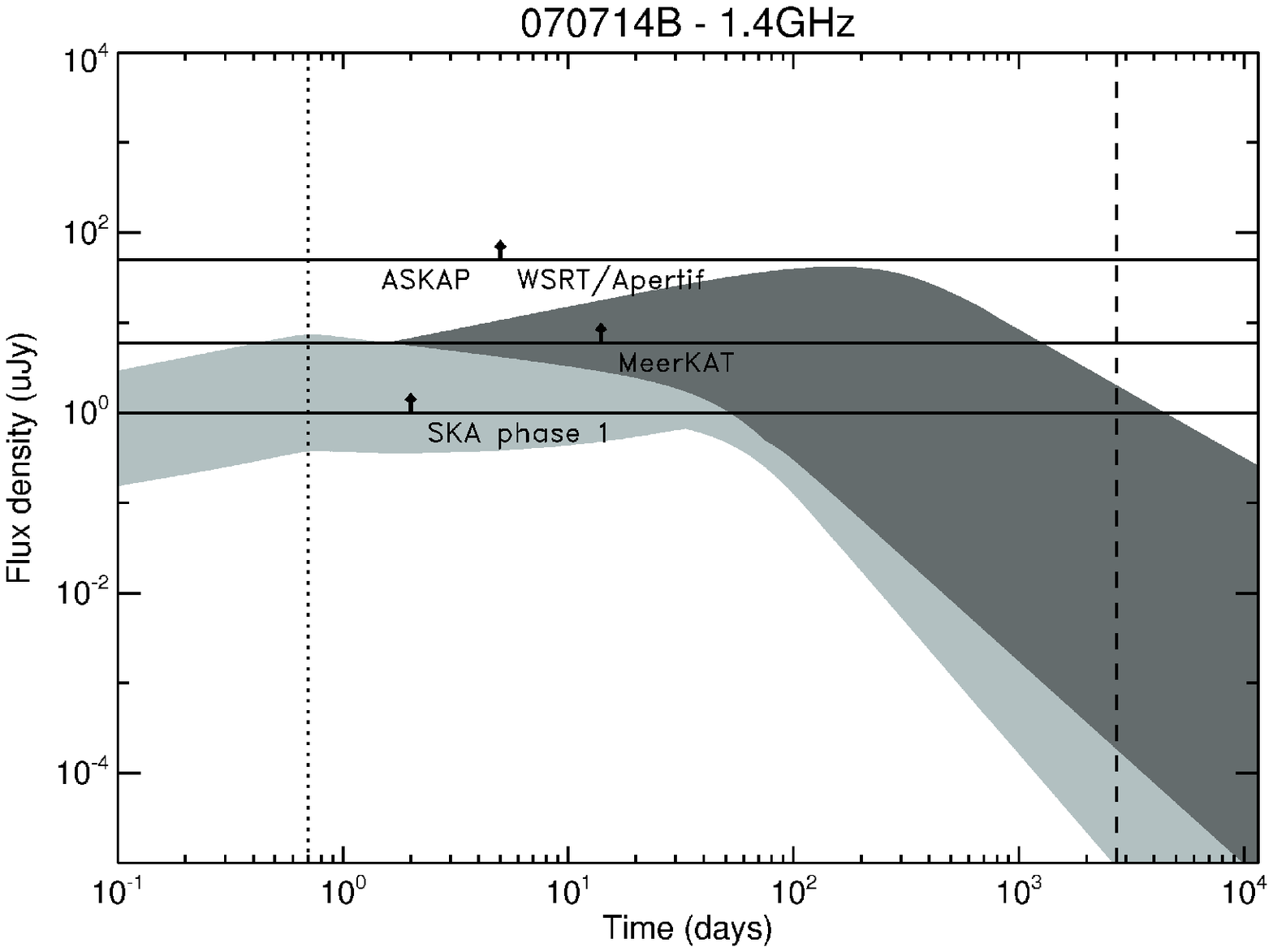}
\includegraphics[width=7.6cm]{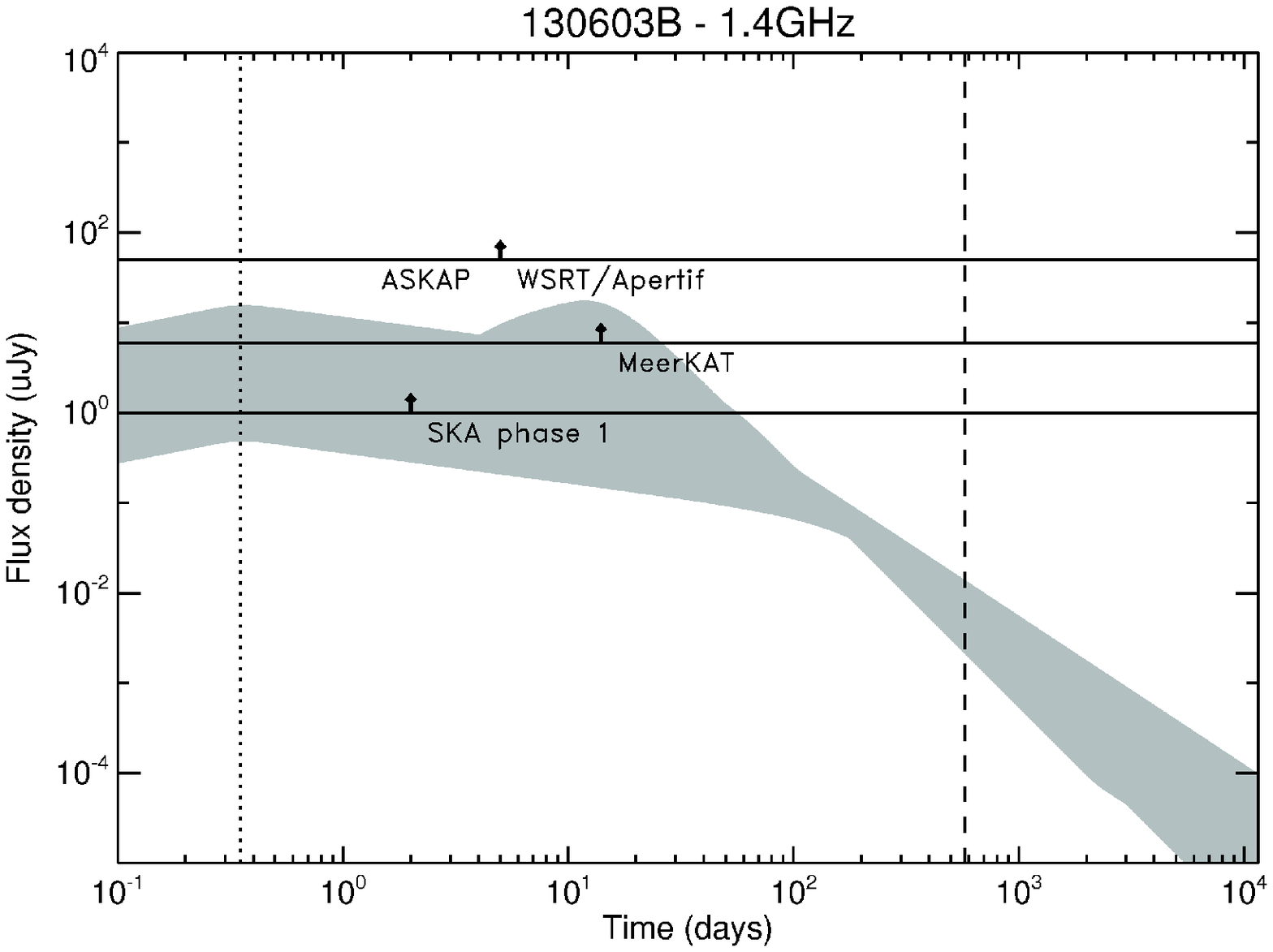}
\includegraphics[width=7.6cm]{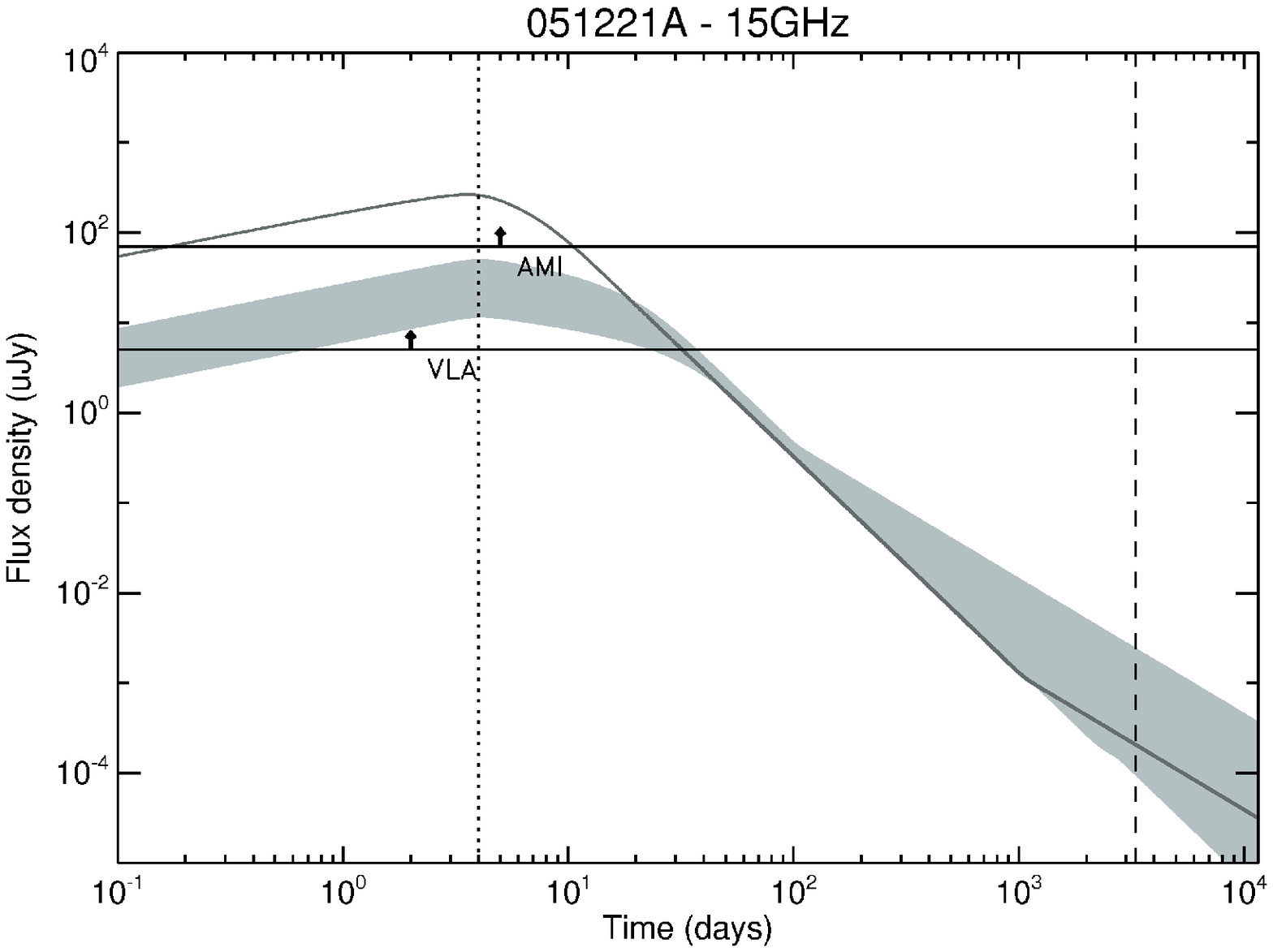}
\includegraphics[width=7.6cm]{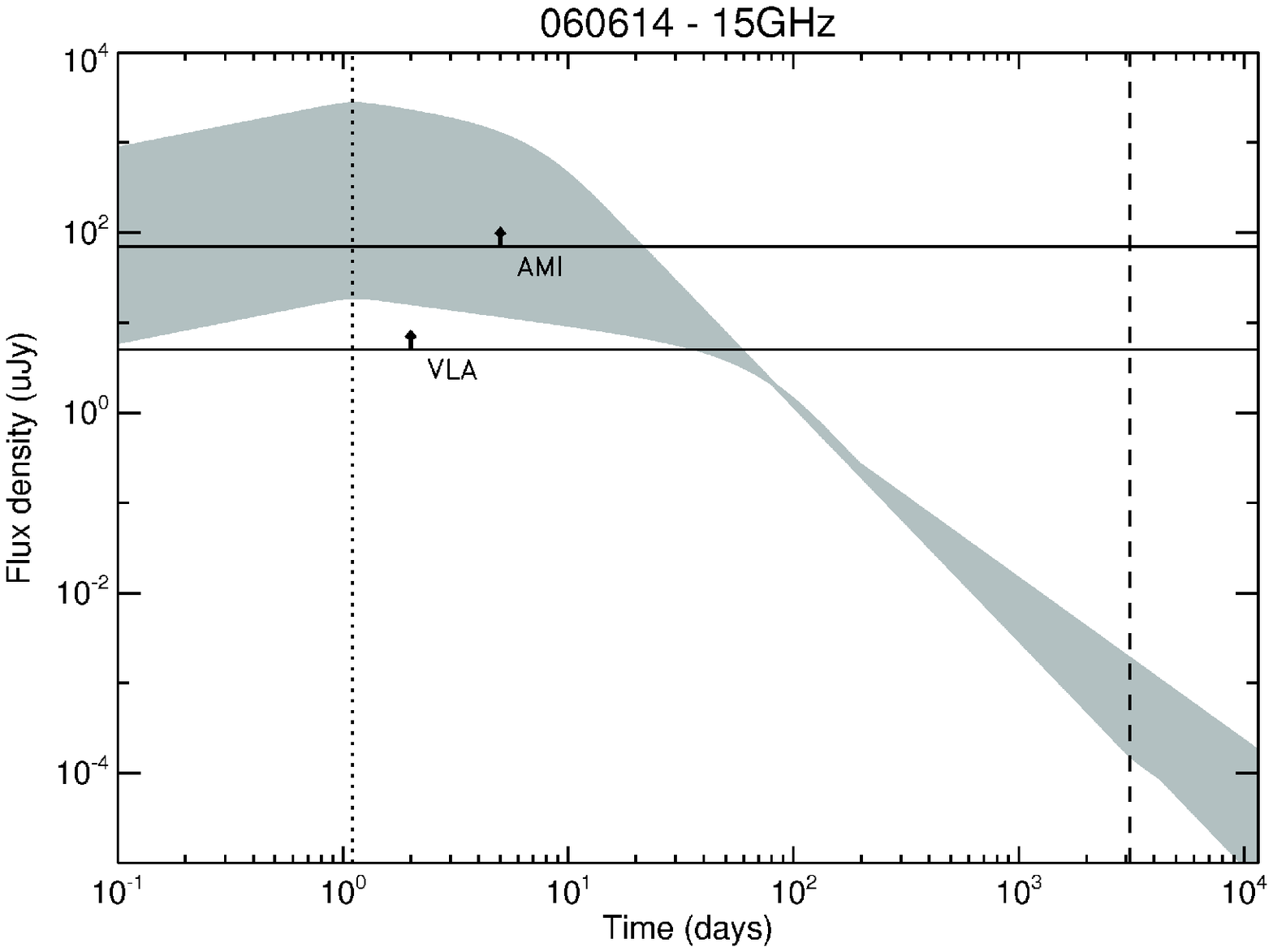}
\includegraphics[width=7.6cm]{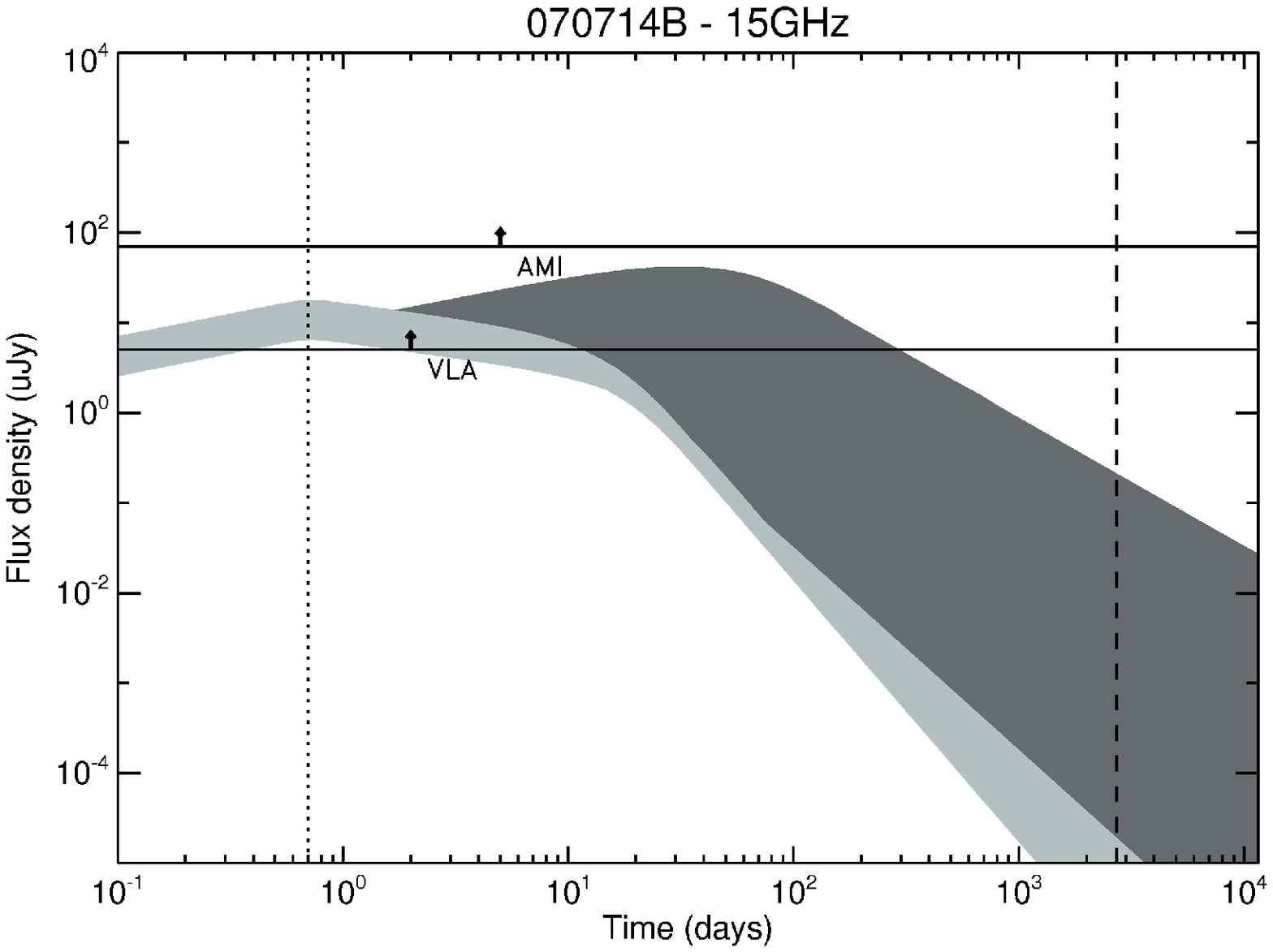}
\includegraphics[width=7.6cm]{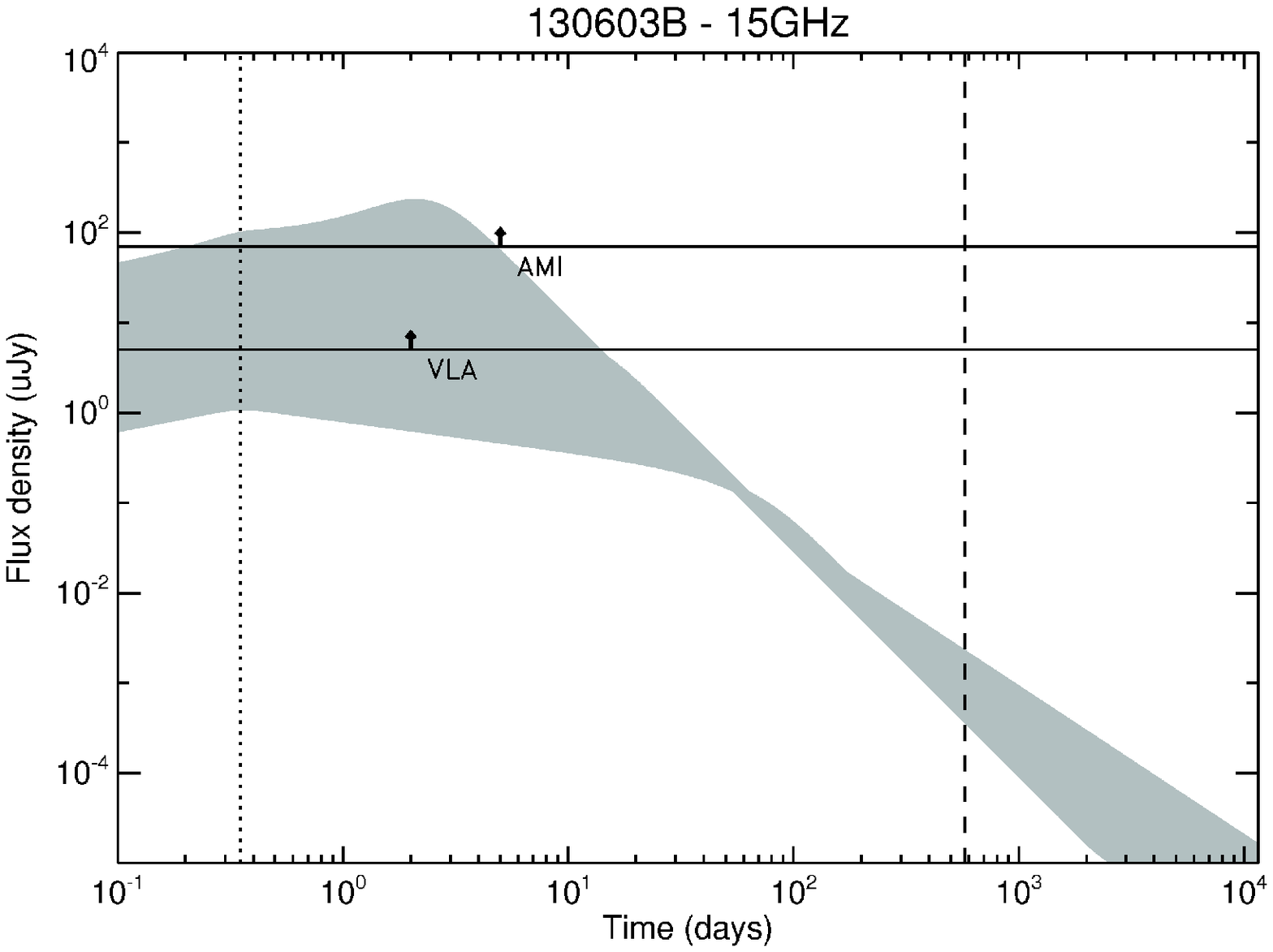}
\end{center}
\caption{{\bf continued.} $1.4$ and $15$~GHz.}
\end{figure*}
\addtocounter{figure}{-1}
\begin{figure*}
\begin{center}
\includegraphics[width=7.6cm]{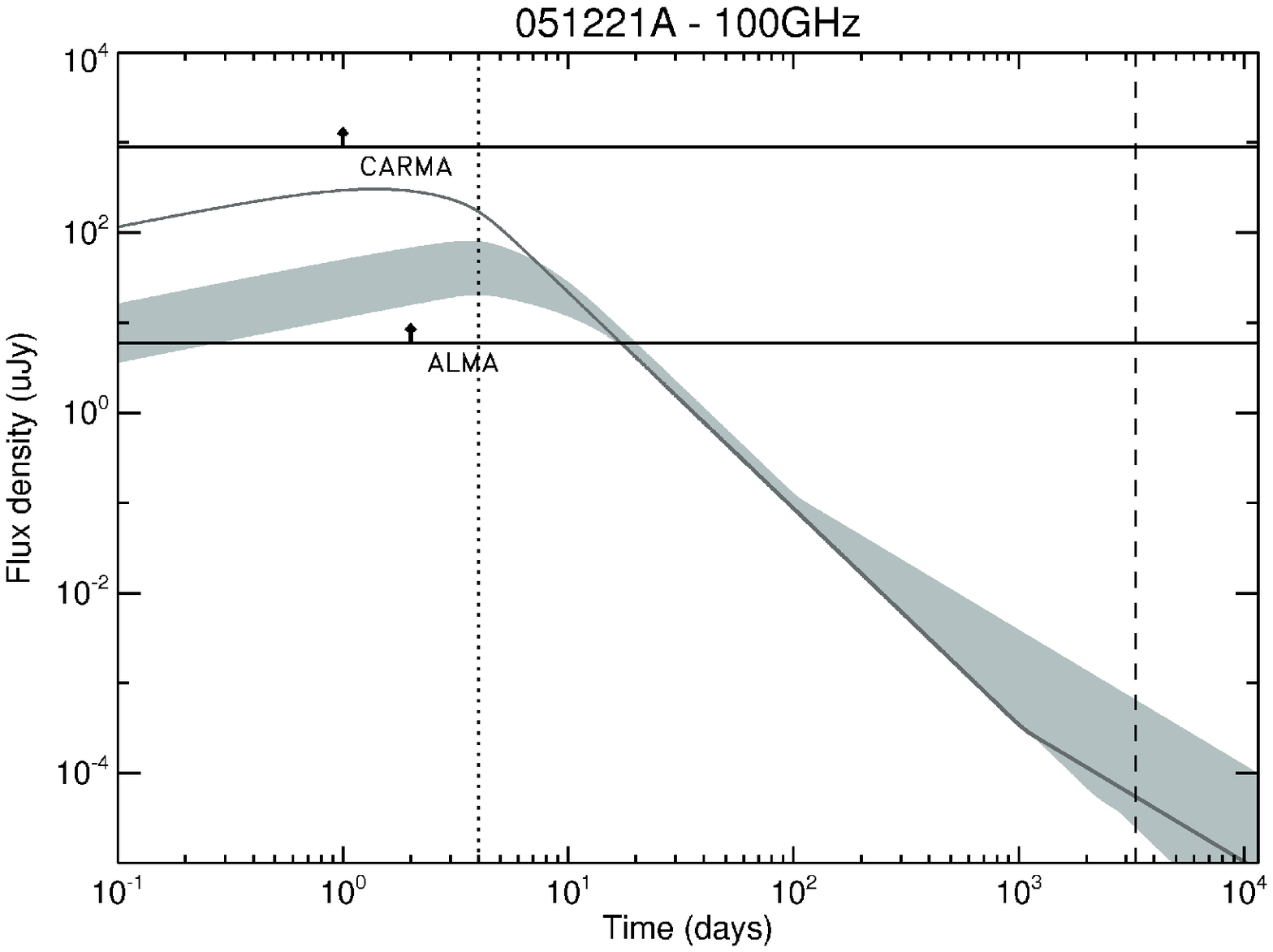}
\includegraphics[width=7.6cm]{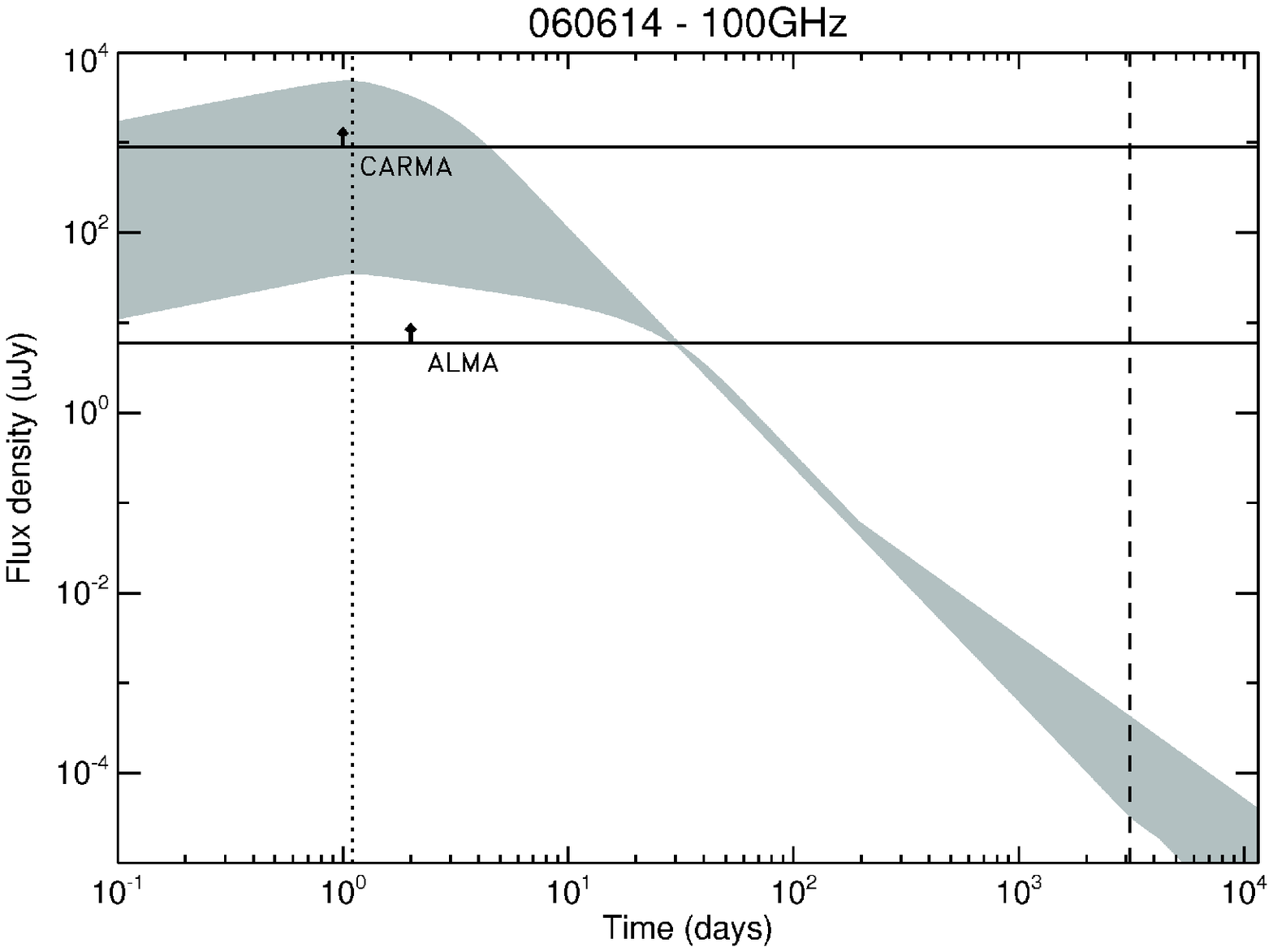}
\includegraphics[width=7.6cm]{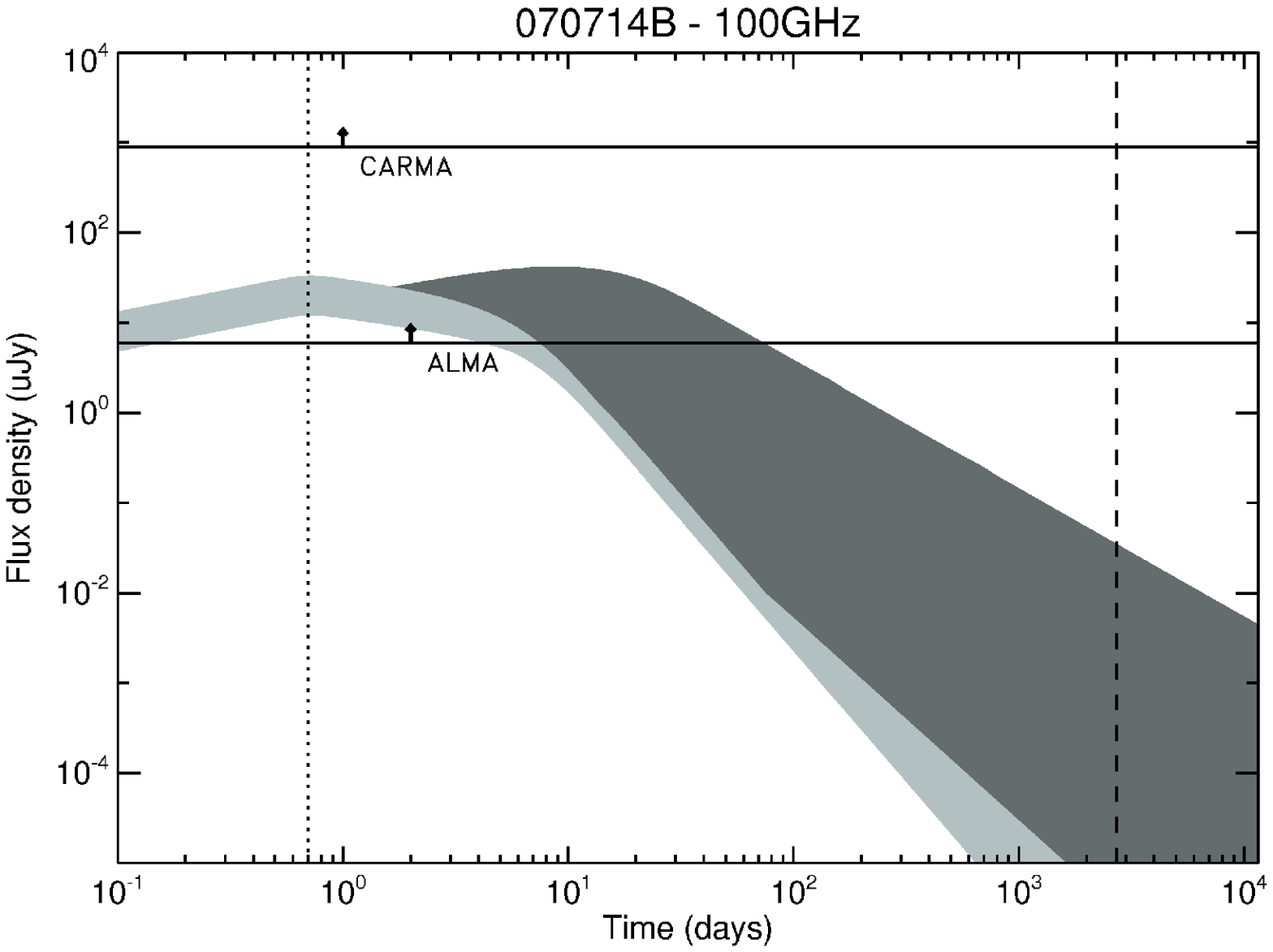}
\includegraphics[width=7.6cm]{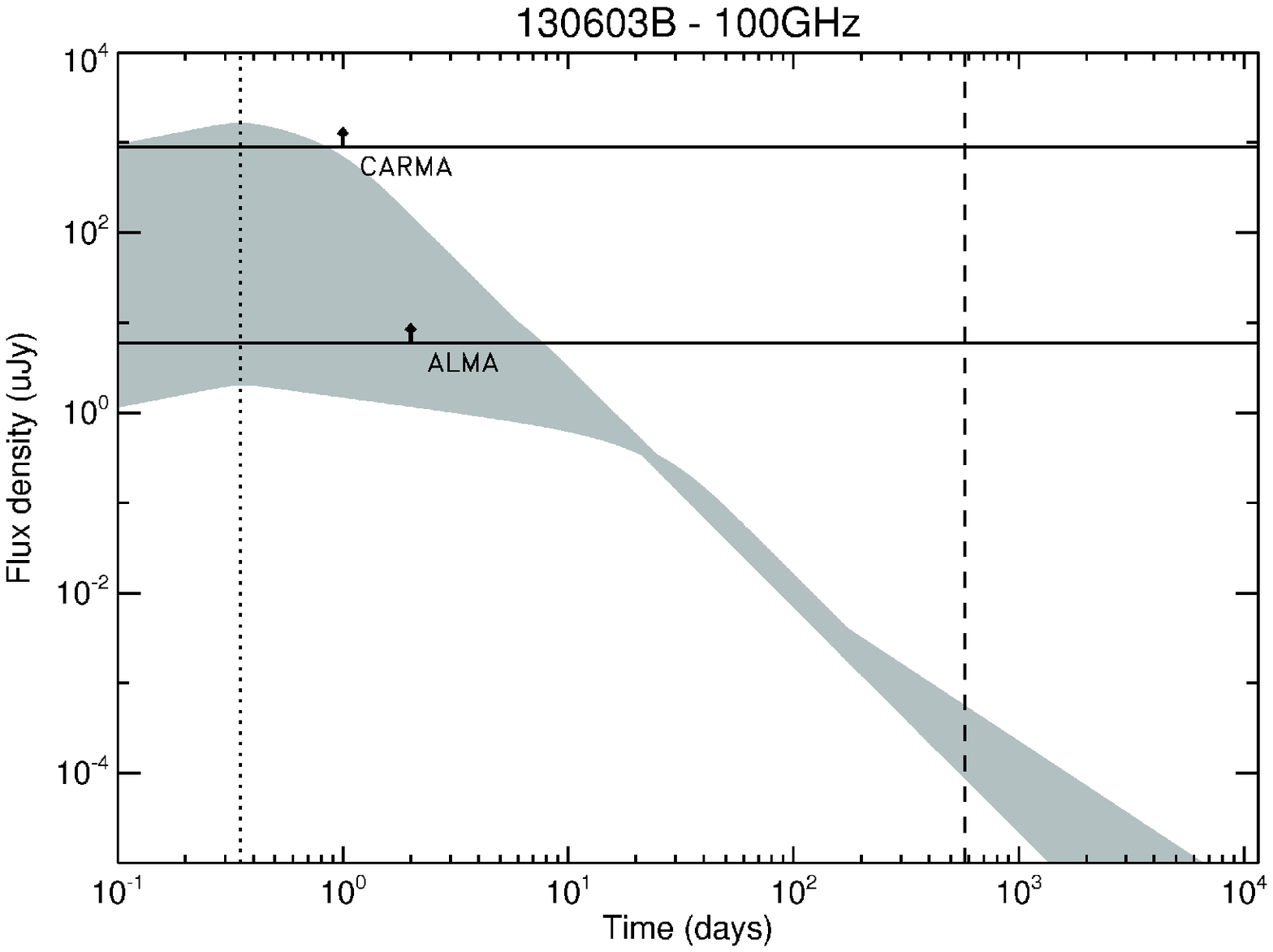}
\end{center}
\caption{{\bf continued.} $100$~GHz.}
\end{figure*}

The models that successfully match the available broadband observations in Section~\ref{sec:results} are used to create synthetic light curves in a variety of radio frequencies: $60$ and $150$~MHz, and $1.4$, $15$, and $100$~GHz. The light curves combine to give a region of predicted flux densities, showing the bounds of what the radio afterglow should have looked like for each GRB at each frequency, given the imposed restrictions of our specific physical model. This is plotted in Fig.~\ref{fig:results}. Table~\ref{tab:limits} shows the sensitivity thresholds for modern-day and future radio telescopes that observe at the frequencies plotted, and a selection of these are superimposed on the light curves. We assess the detectability of each GRB radio afterglow. The flux densities are in general modest, typically peaking in the $\mu$Jy range; however, the results for the anomalously bright GRB 060614 do extend up to mJy. The signal from each GRB is suppressed by the jet break, which curtails the initial brightening of the emission early on in the light curve in most cases. The region either side of this break usually represents the best opportunity to observe the radio afterglow.

At the lower frequencies ($60$ and $150$~MHz), only the Low Frequency Array (LOFAR) at $150$~MHz gets close to being within an order of magnitude of our predictions. The picture is slightly better moving to higher frequencies; in the near future at $1.4$~GHz, the Westerbork Synthesis Radio Telescope (WSRT)/Apertif and the Australian Square Kilometre Array Pathfinder (ASKAP) will be sensitive enough to be capable of observing the brighter models in GRB 060614, and graze the upper limits of the GRB 070714B predictions. MeerKAT would have been capable of detecting at least the upper portion of all four bursts, and could have resolved the entire predicted region of GRB 060614 if observations had been made around the time of the jet break.

At $15$~GHz, the Arcminute Microkelvin Imager (AMI) is capable of observing the upper reaches of the predictions for all but the highest $z$ burst (GRB 070714B) for around a week, possibly even a month for the brighter portion of GRB 060614. The Very Large Array (VLA), in its expanded capacity \citep{Perley11}, would have been able to go deeper than our lower limits in each burst except GRB 130603B in the first week, and provide meaningful limits on the evolution of the radio afterglow for up to a year after trigger. Finally, at $100$~GHz the Combined Array for Research in Millimetre-wave Astronomy (CARMA) may have been able to detect the brightest models in GRB 060614 and GRB 130603B, and the Atacama Large Millimetre Array (ALMA) would have been able to provide limits similar to those mentioned for the VLA, with a window of weeks in GRB 051221A and GRB 070714B, and months in GRB 060614, where the entire predicted region lay above its sensitivity threshold.

Our model fluxes show that previous radio observations, while able to limit some of the physical parameter space, were not deep enough to place serious constraints on the magnetar model. However, the recently upgraded VLA \citep{Perley11} and ALMA are now at $\mu$Jy sensitivity, deep enough to probe even the faintest predicted models. Either telescope can now provide meaningful and highly constraining restrictions on a central engine invoking dipole spin-down injection into a forward shock by making observations within the first week or two after trigger, assuming the four GRBs discussed here are representative of the sample as a whole. Since our sample contains the highest recorded spectroscopic SGRB redshift ($z = 0.9224$; GRB 070714B) and the results in Table~\ref{tab:results} show ISM densities at or near the observed lower limit $n_0 \sim 10^{-5}$~cm$^{-3}$, we suggest that our sample does represent SGRB and EE GRB radio fluxes as a whole, rather than the most luminous cases.

The Square Kilometre Array (SKA) paints a rather brighter picture for the future; our results suggest that even at phase 1, we should expect to see magnetar-injection driven $1.4$~GHz afterglows for months after trigger if the model is to be believed. All four GRBs shown here would be observable for months, in some cases up to a year after trigger, with only the very faintest models in GRB 070714B and GRB 130603B lying below the sensitivity threshold. By phase 2, all four of the radio afterglows in our sample would have been visible for a year or more, and the entire predicted flux density region could be explored for each with the correct observing strategy. Our findings are in agreement with \citet{Feng14}, who simulated radio afterglow light curves for compact object mergers at the advanced Laser Interferometer Gravitational-wave Observatory (aLIGO) horizon. We consider here the simplest case of merger followed by injection; however, the radio signal from these mergers may be further enhanced by other processes such as macronovae \citep{Piran13}.

\section{Conclusions}\label{sec:conclusions}

We have performed order of magnitude fitting to the broadband afterglows of a sample of four GRBs. We use a physically motivated central engine, invoking energy injection into a forward shock from a magnetar as it rapidly loses angular momentum along open magnetic field lines. By imposing the limitations of a self-consistent central engine for the energy profile of each GRB, we are able to narrow the available parameter space for the physics underlying the evolution of the blast wave as it expands into the ambient medium. Combinations of these parameters are tested against the data, resulting in a family of models that accurately recreate observations. These models are then used to predict the radio signature from the central engine, and are assessed for detectability.

Our results show that current broadband observations are consistent with the magnetar injection model, as we find physical parameters that lie within the allowed ranges for all bursts. Some discrepancies exist at radio frequencies, suggesting that previous early detections captured emission from a reverse shock propagating backwards through the ejecta, rather than a forward shock moving outwards into the ISM. We find that while recent observational detection thresholds are not constraining to the magnetar model, state-of-the-art facilities such as the upgraded VLA and ALMA are now capable of observing to depths greater than our predicted flux density range if observations are made in the first few weeks, and to maximum sensitivity. We also show that SKA will be capable of observing to depths in excess of our model predictions, and hence is expected to observe these signatures, or impose strict limits on the physical parameters.

\section{Acknowledgements}

We thank Hendrik van Eerten for helpful feedback that improved the manuscript. We also thank the anonymous referee for useful comments that aided the clarity of this paper. BG acknowledges funding from the Science and Technology Funding Council. AJvdH acknowledges support from the European Research Council via Advanced Investigator Grant no. 247295 (PI: R.A.M.J. Wijers). The work makes use of data supplied by the UK \emph{Swift} Science Data Centre at the University of Leicester and the \emph{Swift} satellite. \emph{Swift}, launched in November 2004, is a NASA mission in partnership with the Italian Space Agency and the UK Space Agency. \emph{Swift} is managed by NASA Goddard. Penn State University controls science and flight operations from the Mission Operations Center in University Park, Pennsylvania. Los Alamos National Laboratory provides gamma-ray imaging analysis.

\bibliographystyle{mn2e}
\bibliography{/local/data/72/bpg6/papers/ref}

\label{lastpage}

\newpage

\appendix
\section{Flux density equations}\label{sec:equations}
Using the equations below, the flux ($F$; erg~cm$^{-2}$~s$^{-1}$) observed in a bandpass bounded by a lower limit $\nu_l$ and upper limit $\nu_h$ (both Hz) can be converted to a flux density ($F_{\nu_{\rm p}}$; Jy) at the bandpass logarithmic mid-point ($\nu_{\rm p}$; Hz), assuming a power-law spectrum with an index $\beta$ ($F_{\nu} = \nu^{-\beta}$).

\begin{align}
F_{\nu_{\rm p}}=& \hspace{0.1cm} \frac{(\beta-1)F}{\nu_l}\bigg(\frac{\nu_l}{\nu_h}\bigg)^{\beta/2}\bigg[1-\bigg(\frac{\nu_h}{\nu_l}\bigg)^{1-\beta}\bigg]^{-1} \quad
&\text{for }\beta > 1 \nonumber \\
F_{\nu_{\rm p}}=& \hspace{0.1cm} \frac{F}{\nu_{\rm p}}\bigg[{\rm ln}\bigg(\frac{\nu_h}{\nu_l}\bigg)\bigg]^{-1} \quad
&\text{for }\beta = 1 \nonumber \\
F_{\nu_{\rm p}}=& \hspace{0.1cm} \frac{(1-\beta)F}{\nu_h}\bigg(\frac{\nu_h}{\nu_l}\bigg)^{\beta/2}\bigg[1-\bigg(\frac{\nu_l}{\nu_h}\bigg)^{1-\beta}\bigg]^{-1} \quad
&\text{for }\beta < 1
\end{align}

\end{document}